\documentclass[12pt,nofootinbib]{revtex4}
\usepackage{amssymb}
\usepackage{amsmath, graphicx}
\usepackage{slashed}
\usepackage[unicode]{hyperref}
\usepackage{csquotes}
\usepackage{braket}
\usepackage{natbib}
\usepackage{array}
\usepackage{xcolor}
\usepackage{multirow}

\begin{document}
\title{Correlation between gravitational wave frequencies as a signature of new black hole physics}
\author{Ngo Phuc Duc Loc}
\thanks{Email: locngo148@gmail.com}

\affiliation{Physics Division, National Center for Theoretical Sciences,
National Taiwan University, Taipei 106319, Taiwan}

\begin{abstract}
We point out that for black holes (BH) of the same mass, despite of arising from two entirely different physical arguments, the peak frequency of the coarse-grained gravitational wave (GW) spectrum of gravitons $f_\text{peak}^\text{graviton}$ is very close to the peak frequency of the GW spectrum produced from binary BH mergers $f_\text{peak}^\text{merger}$. In particular, we find that $f_\text{peak}^\text{graviton}/f_\text{peak}^\text{merger}\approx 2.3$ independent from BH mass or population. If BHs are still evaporating, the peak frequencies scale as $\propto M_i^{-1}$ with $M_i$ is the initial BH mass. If BHs evaporated in the radiation (matter) dominated era, the peak frequencies scale as $\propto M_i^{1/2}$ ($\propto M_i$) respectively after taking into account redshift effect.  Any deviation from the aforementioned magical ratio would indicate new BH physics. The sweetest frequency ballpark to look for this correlation is around $10^{14}\ \rm Hz-10^{20}\ \rm Hz$ where both GW spectra are allowed to have maximum amplitude. As a demonstration of what the new physics could be, we consider a phenomenological self-similar regular BH model of Hayward featuring a single hair parameter. The observed ratio of peak frequencies can be used to directly infer the hair parameter, along with crucial properties of the novel BH solution such as Hawking temperature, horizon size, and innermost stable circular orbit. We also comment on the pros and cons of correlating frequencies among other GW sources such as formation, population fluctuation, and sudden transition of adiabatic perturbation. 
\end{abstract}

\maketitle

\tableofcontents

\section{Introduction}
Motivated from the need to cure the singularity problem of the usual black hole (BH) solutions, many regular black hole (RBH) models have been proposed, ranging from early phenomenological models \cite{Bardeen,Hayward:2005gi,Dymnikova:1992ux,Culetu:2013fsa,Culetu:2014lca,Ghosh:2014pba,Simpson:2019mud,Simpson:2018tsi} to quantum gravity-inspired metrics \cite{Peltola:2008pa,Peltola:2009jm,Bianchi:2018mml,DAmbrosio:2018wgv,Zhang:2024khj}. Recent progresses in the study of dynamical formation of RBH \cite{Bueno:2024dgm,Bueno:2024eig,Bueno:2024zsx,Bueno:2025zaj,Vertogradov:2025snh} also marked an important step toward a more complete picture.

In the context of cosmology, recent studies of regular primordial black holes (RPBHs) have received great interests. RPBHs as cosmic expansion accelerators has been studied in \cite{Dialektopoulos:2025fzp,Dialektopoulos:2025mfz}. Cosmologically coupled RPBHs has been studied in \cite{Cadoni:2023lum,Farrah:2023opk,Croker:2024jfg,Faraoni:2024ghi,Calza:2024qxn}. For non-self-similar RPBHs, their remnant states could be a dark matter (DM) candidate  \cite{Davies:2024ysj,Trivedi:2025vry,Ong:2024dnr,Dymnikova:2015yma} though this possibility also received some skepticism \cite{Khodadi:2025icd}. For self-similar RPBHs, they evaporate completely in a runaway manner similar to the singular Schwarzschild BH but the evaporation timescale is different and determined by the choice of the regularizing parameter (also known as BH's ``hair"). Heavy self-similar RPBHs as a DM candidate has been explored in \cite{Calza:2024fzo,Calza:2024xdh,Calza:2025mwn}. Particle DM can also be produced from evaporation of ultra-light self-similar RPBHs \cite{Loc:2025mzc} and isocuvature gravitational wave (GW) constraint of such scenario has been studied in \cite{Loc:2025whx}.

Given the rising interests in the idea of RBH, it is natural to ask how we can observationally discriminate between singular and regular BHs or exotic compact objects in general. A variety of methods have been proposed such as studying shadows of BH  \cite{Vagnozzi:2022moj}, ringdown phase and quasinormal modes  \cite{Flachi:2012nv,Pedrotti:2024znu,Cardoso:2016rao,Rosato:2026apq,Lu:2025vol}, spin-induced quadrupole moment \cite{Krishnendu:2025rud}, tidal love number  \cite{Wang:2025oek,Andres-Carcasona:2025bni},  echoes and light rings \cite{Rao:2025rop}, accretion \cite{Akbarieh:2023kjv,Boshkayev:2023fft}. These attempts only considered specific isolated events. In this paper, we propose a new way to detect novel BH physics by correlating peak frequencies of different stochastic GW backgrounds produced by a population of BHs. This method is most promising from the observational standpoint for small-mass BHs that have primordial origin, though its theoretical logic could be applied to any BHs in general. Self-similar ultra-light PBHs could generate a sufficiently strong GW background from evaporation of gravitons \cite{Page:1976df,Page:1976ki,Dong:2015yjs,Ireland:2023avg,Anantua:2008am,Dolgov:2011cq}. They can also generate a GW background from binary mergers \cite{Phinney:2001di,Sasaki:2018dmp,Wang:2019kaf,Sasaki:2016jop,Ajith:2007kx,Zhu:2011bd,Ali-Haimoud:2017rtz,Raidal:2024bmm,Raidal:2017mfl,LIGOScientific:2025kry}. We show that the correlation between the peak frequencies of these two different GW production mechanisms is capable of distinguishing singular and regular PBH, or more generally detecting new BH physics. We will also discuss advantages and disadvantages of correlating peak frequencies with other GW sources.

The rest of this paper is organized as follows. We first introduce the parameterization of new BH physics in Sec. \ref{sec:parameterization of new BH physics}. Next, we provide general formulae for graviton frequency, merger frequency, their correlation, effects of cosmological redshift, and recommend an optimal frequency ballpark to test the scenario in Sec. \ref{sec:general formulae}. We then demonstrate the idea with an example of the Hayward metric in Sec. \ref{sec:example}. Comments on the pros and cons of correlating peak frequencies of other GW sources are given in Sec.\ref{sec: comments}. Conclusions are in Sec. \ref{sec:conclusion}. Derivation of the innermost stable circular orbit for a general static, spherically symmetric metric is provided in Appendix \ref{appendix: rISCO}, whereas some calculations of GW from gravitons are given in Appendix \ref{Appendix: graviton}. We use the natural units in which $\hbar=c=k_\text{B}=1$.

\section{Parameterization of new black hole physics}\label{sec:parameterization of new BH physics}

We first give a brief review of BH evaporation \cite{Loc:2025mzc}, focusing on the aspects that will be most relevant for our present purpose. We consider a class of static, spherically symmetric metric of the form
\begin{equation}
    ds^2=-f(r,l)dt^2+\frac{dr^2}{g(r,l)}+h(r,l)d\Omega^2,
\end{equation}
where $l\geq0$ is the dimensionless regularizing parameter and $d\Omega^2$ is the usual 2-sphere metric\footnote{While both $r$ and $l$ appear inside the bracket, it should be understood that $r$ is a variable whereas $l$ is a parameter.}. The Hawking temperature is given by \cite{Hawking:1974rv,Hawking:1975vcx,Gibbons:1977mu}
\begin{equation}
    T_\text{H}=\frac{f'(r,l)}{4\pi}\sqrt{\frac{g(r,l)}{f(r,l)}}\Bigg|_{r=r_\text{H}},
\end{equation}
where prime denotes derivative with respect to $r$, $r_\text{H}$ is the outer horizon radius and is the largest root of the equation $f(r,l)=0$. We find it useful to define the ratios
\begin{equation}\label{eq: A and B}
    A(l)\equiv\frac{T_\text{H}}{T_\text{H,Sch}}\hspace{0.5cm},\hspace{0.5cm}
    B(l)\equiv\frac{R_\text{H}}{R_\text{H,Sch}}\hspace{0.5cm},\hspace{0.5cm}C(l)\equiv\frac{r_\text{ISCO}}{r_\text{ISCO,Sch}}.
\end{equation}
Here, $R_\text{H}$ is the physical horizon radius and is related to the coordinate horizon radius $r_\text{H}$ via $R_\text{H}=\sqrt{h(r_\text{H},l)}$. For metrics with standard angular metric function $h(r,l)=r^2$, such as the usual Schwarzschild BH, they are identical but in general they could be different. $r_\text{ISCO}$ is the radius of the innermost stable circular orbit (ISCO). See Appendix \ref{appendix: rISCO} for details of how to obtain $r_\text{ISCO}$ for a general metric. Quantities without the subscript ‘Sch’ correspond to the novel BH solution, while those carrying the subscript ‘Sch’ refer to the Schwarzschild BH and are given by:
\begin{equation}\label{eq: T_Sch and r_Sch}
    T_\text{H,Sch}=\frac{1}{8\pi GM}\hspace{0,5cm},\hspace{0.5cm}r_\text{H,Sch}=2GM\hspace{0.5cm},\hspace{0.5cm}r_\text{ISCO,Sch}=6GM
\end{equation}
where $G$ is the Newtonian gravitational constant. In natural units, we have $G=1/m_\text{pl}^2$ with $m_\text{pl}$ is the Planck mass. The novel metrics are constructed such that they can reduce to the standard Schwarzschild BH case in the limit $l\rightarrow 0$, so that $A(l\rightarrow 0)=B(l\rightarrow 0)=C(l\rightarrow 0)=1$. While this parameterization was first introduced in \cite{Loc:2025mzc,Loc:2025whx} to characterize regular BH, the framework can be applied to any novel BH solution in general.

In the normalized form, the Hawking temperature is
\begin{equation}\label{eq:T_H}
    \left(\frac{T_\text{H,i}}{\rm GeV}\right)=1.06\times 10^{13}A(l)\left(\frac{\rm g}{M_i}\right).
\end{equation}
The evolution of BH mass is
\begin{equation}\label{eq:M(t)}
    M(t)=M_i\left(1-\frac{t-t_i}{\tau}\right)^{1/3},
\end{equation}
where $M_i$ is the initial mass:
\begin{equation}
    \left(\frac{M_i}{\rm g}\right)\simeq 9.45\times 10^{31}\gamma \left(\frac{106.75}{g_{*,i}}\right)^{1/2}\left(\frac{\rm GeV}{T_i}\right)^2
\end{equation}
with $\gamma$ is fraction of horizon mass at the formation time\footnote{Typically, $\gamma\simeq 0.2$ for Schwarzschild BH but is kept here as a general parameter. The peak frequencies of GWs from mergers and gravitons, and thus their correlations, are independent from this piece of information.},  $T_i$ is the cosmic temperature at formation time, and $t_i$ is the formation time given by
\begin{equation}\label{eq:ti}
    \left(\frac{t_i}{\rm sec}\right)=2.48\times 10^{-39}\gamma^{-1}\left(\frac{M_i}{\rm g}\right),
\end{equation}
and $\tau$ is the lifetime of BH:
\begin{equation}\label{eq:tau}
    \left(\frac{\tau}{\rm sec}\right)=1.57\times 10^{-27}\frac{1}{B(l)^2A(l)^4}\left(\frac{106.75}{g_*(T_\text{H})}\right)\left(\frac{M_i}{\rm g}\right)^3.
\end{equation}
The cosmic temperature at which the PBH evaporates can be found from the condition $\Gamma_\text{BH}\sim H_\text{eva}$ with $\Gamma_\text{BH}$ is the BH's decay rate and $H_\text{eva}\propto T_\text{eva}^2$ is the usual Hubble rate, which gives
\begin{equation}\label{eq:T_eva}
    \left(\frac{T_\text{eva}    }{\rm GeV}\right)=1.73\times 10^{10}A(l)^2B(l)\left(\frac{g_*(T_\text{H})}{106.75}\right)^{1/2}\left(\frac{106.75}{g_{*,\rm eva}}\right)^{1/4}\left(\frac{\rm g}{M_i}\right)^{3/2}.
\end{equation}
As entropy is conserved from after PBH evaporation to today, the scale factor at evaporation time is
\begin{equation}\label{eq:a_eva}
    a_\text{eva}=\left(\frac{g_\text{s,0}}{g_\text{s,eva}}\right)^{1/3}\left(\frac{T_0}{T_\text{eva}}\right)=4.43\times 10^{-24}\frac{1}{A(l)^2B(l)}\left(\frac{106.75}{g_{*,\rm eva}}\right)^{1/12}\left(\frac{106.75}{g_*(T_\text{H})}\right)^{1/2}\left(\frac{M_i}{\rm g}\right)^{3/2}.
\end{equation}
In the above equations, $g_*\sim \mathcal{O}(100)$ factors are relativistic degrees of freedom at the corresponding temperature, the subscript ``eva" and ``0" denotes quantities at evaporation time and today respectively, $g_\text{s}$ factors are entropy degrees of freedom and at high temperature we have $g_\text{s,eva}\sim g_{*,\rm eva}$, and we also used the fact that $g_\text{s,0}=3.94$ and $T_0=2.3\times 10^{-13}\ \rm GeV$.

\section{General formulae}\label{sec:general formulae}

In this section, we provide general formulae for peak frequencies of GW spectrum from gravitons and binary BH mergers in the source frame and establish their correlation. We then discuss how the apparent dependence of peak frequencies on BH mass changes when the sources are at cosmological distance in different eras. We finally point out the sweetest frequency ballpark for future GW detectors to look for  this correlation and hence search for new BH physics.

\subsection{Graviton frequency}

The most well-motivated massless spin-2 particle candidate that could be produced from BH's evaporation is known as \textit{``graviton"}, which is assumed to be the excitation of the gravitational field if gravity can be quantized like other forces in nature. Intuitively, we can already expect that the graviton frequency at production time should be of order of Hawking temperature $\omega_\text{eva}\sim T_\text{H}$. A more formal treatment gives a correction factor of order $\sim O(1)$ as discussed in Appendix \ref{Appendix: graviton}. We quote the result here:
\begin{equation}\label{eq:fpeak_graviton_source frame}
    \frac{f_\text{peak}^\text{graviton}}{\rm Hz}\simeq 10^{37}A(l)\left(\frac{\rm g}{M_i}\right).
\end{equation}
This is the peak frequency of GW from gravitons in the source frame.

\subsection{Merger frequency}

The standard way to estimate the peak frequency of GW spectrum from binary BHs merger is to note that the inspiralling phase ends when the two BHs are closer than $r_\text{ISCO}$. The angular velocity at this distance is $\Omega=\sqrt{GM/r_\text{ISCO}^3}$ with $M$ is the total mass of the binary, so the corresponding angular frequency is $f_\text{ISCO}=\Omega/2\pi$. The associated GW frequency in the source frame is then twice this value\footnote{The gravitational wave frequency is twice the orbital angular frequency because GWs arise from the quadrupole moment, which oscillates at double the orbital rate.}, which gives
\begin{equation}
    f_\text{GW,peak}^\text{merger}=\frac{\Omega}{\pi}=\frac{1}{\pi}\sqrt{\frac{GM}{r_\text{ISCO}^3}}=\frac{1}{\pi}\sqrt{\frac{GM}{C(l)^3(r_\text{ISCO,Sch})^3}}\xrightarrow{\text{SI units}}\frac{1}{C(l)^{3/2}}\frac{c^3}{\pi\ 6^{3/2}GM},
\end{equation}
where we recovered the standard units in the last step. Assuming equal-mass BHs and putting in numerical factors, we get
\begin{equation}\label{eq:f peak merger}
    \frac{f_\text{GW,peak}^\text{merger}}{\rm Hz}\simeq \frac{4.4\times 10^{36}}{C(l)^{3/2}}\left(\frac{\rm g}{M_i}\right).
\end{equation}
In the limit of no new physics with $C(l=0)=1$, two BHs with mass $\sim 30M_\odot\sim3\times 10^{34}\rm g$ gives $f_\text{GW,peak}^\text{merger}\sim 150\ \rm Hz$ consistent with the GW150914 event \cite{LIGOScientific:2016aoc,LIGOScientific:2016wyt}. However, it is still possible to accommodate a novel BH solution. For example, the maximum possible deviation of the Hayward metric from the Schwarzschild metric gives $C_\text{Hayward}(l\approx0.77)\approx 0.93$ which leaves little difference on the maximum frequency.

Even if astrophysical BHs are standard physics, that does not automatically rule out potentially novel BH physics that could be present in the primordial Universe. In fact, while our proposed method could in principle be applied to any BHs, the most promising case from the observational standpoint is ultra-light PBHs that evaporated completely within the first 1 second after the Big Bang (see Sec. \ref{subsec:where to look?}).

\subsection{The correlation}

If we observe only one of the peak frequencies, there will be a degeneracy from the mass and the hair parameter. To break this degeneracy, we need to observe both peaks. Unlike other potential primordial sources of GW, the correlation between graviton frequency and merger frequency genuinely captures the BH physics, independent from the formation mechanism or the somewhat artificial frequency cutoffs of GW sources associated with PBH reheating if there is PBH domination (see Sec. \ref{sec: comments}). From Eqs. \ref{eq:fpeak_graviton_source frame} and \ref{eq:f peak merger}, we can eliminate the mass variable to get
\begin{equation}\label{eq:frequency correlation}
    \frac{f_\text{peak}^\text{graviton}}{f_\text{peak}^\text{merger}}\simeq 2.3\ A(l)C(l)^{3/2}.
\end{equation}
The standard Schwarzschild solution is recovered in the limit $A(l\rightarrow0)=C(l\rightarrow 0)=1$, which gives $f_\text{peak}^\text{graviton}/f_\text{peak}^\text{merger}\simeq 2.3$. Physically, the fact that both peak frequencies scale as $\propto M_i^{-1}$ in the source frame should not be a big surprise. More mass means lower Hawking temperature and thus lower graviton frequency. More mass also means stronger gravity, which makes the binary forms at greater distance and thus lower merger frequency. The surprising thing is that the numerical factors almost cancel out in the Schwarzschild limit, even though one effect is quantum and the other one is classical. Novel BH physics with $A(l)\neq 1$ and/or $C(l)\neq 1$ may break this coincidence.

\subsection{Cosmological redshift}\label{subsec:redshift}
\begin{figure}[h!]
    \centering
    \includegraphics[width=0.8\linewidth]{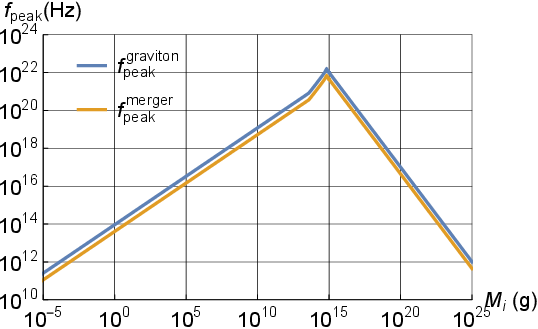}
    \caption{$f_\text{peak}^\text{graviton}$ and $f_\text{peak}^\text{merger}$ as a function of $M_i$ for Schwarzschild BH. The peak frequencies scale as $f_\text{peak}\propto  M_i^{1/2}$ for BH with mass $M_i\lesssim 10^{13}\rm g$ evaporated in RD phase, $f_\text{peak}\propto M_i$ for BH with mass $10^{13}\rm g\lesssim M_i\lesssim 10^{15}\rm g$ evaporated in MD phase, and $f_\text{peak}\propto M_i^{-1}$ for BH with mass $M_i\gtrsim 10^{15}\rm g$ still evaporating today.}
    \label{fig:fpeak_Mi}
\end{figure}

If the sources are at cosmological distance, the frequency is redshifted as $f_0=(a_\text{eva}/a_0)f_\text{eva}$ where $f_0\  (f_\text{eva})$ is the frequency today (at evaporation) and $a_0\ (a_\text{eva})$ is the corresponding scale factor at those moments. The scale factor today could be set to unity $a_0=1$. To determine the redshift factor $a_\text{eva}$, we recall that the age of the Universe at $a_\text{eva}$ is
\begin{equation}
    t_\text{age}^\text{Universe}(a_\text{eva})=\frac{1}{H_0}\int_0^{a_\text{eva}}\frac{da}{a\sqrt{\sum_i\Omega_{i,0}a^{-3(1+w_i)}}},
\end{equation}
where we assumed the standard flat $\Lambda\rm CDM$ model of cosmology with $H_0=67\ \rm km/s/Mpc=2.17\times 10^{-18}s^{-1}$ \cite{Planck:2018vyg} is the Hubble constant today, and $\Omega_{i,0}$ is the fraction of total energy density in specie $i^{\rm th}$ today with equation of state $w_i$. We take numerical values $\Omega_{m,0}=0.32, \Omega_{\rm  rad,0}=8.6\times 10^{-5}, \Omega_{\Lambda,0}=0.68$ \cite{Planck:2018vyg}. Thus, the scale factor at which the BH evaporates can be determined from $\tau_\text{BH}= t_\text{age}^\text{Universe}(a_\text{eva})$ which could be done numerically\footnote{Note that $t_i\ll\tau$ in a typical case and can be ignored \cite{Loc:2025mzc}.}. The resulting redshifted frequency observed today is shown in Fig. \ref{fig:fpeak_Mi}.

The GW spectrum from gravitons receives the most dominant contribution at the time of BH's evaporation and thus peaks at $a_\text{eva}$. Also, in the standard scenario when the merger rate does not increase sharply with redshift, the dominant contribution of GW from binary BH mergers must come from sources at later time because the amplitude of GW is suppressed more for sources at higher redshift \cite{LIGOScientific:2025kry}. Therefore, both GW spectra from gravitons and mergers peak at the time of BH's evaporation since that is the latest time for BHs to still exist, so that the peak frequencies experience the same amount of subsequent redshift (see also \cite{Inomata:2020lmk}). Thus, the ratio $f_\text{peak}^\text{graviton}/f_\text{peak}^\text{merger}\simeq 2.3$ in the source frame is frozen across cosmological history, but the dependence on BH mass of individual components will change depending on which cosmic era the BH evaporated or is still evaporating, as we will see in the following.

To gain further analytical insights into individual peak frequencies, we can consider certain limits as follows:
\begin{itemize}
    \item If the BH evaporated in a radiation dominated (RD) Universe when $a_\text{eva}< a_\text{eq}$, with $a_\text{eq}=3\times10^{-4}$ is the scale factor at matter-radiation equality, we only keep the most dominant term associated with $\Omega_\text{rad}$ to get \footnote{Note that this is consistent with Eq. \ref{eq:a_eva} derived using the entropy conservation argument up to an $\mathcal{O}(1)$ factor.}
    \begin{equation}
        a_\text{eva}\simeq 7.95\times 10^{-24}\frac{1}{B(l)A(l)^2}\left(\frac{M_i}{\rm g}\right)^{3/2}.
    \end{equation}
    This gives the peak frequencies:
    \begin{equation}\label{eq:fpeak_graviton_RD}
        \frac{f_\text{peak}^\text{graviton}}{\rm Hz}\simeq 7.95\times 10^{13}\frac{1}{B(l)A(l)}\left(\frac{M_i}{\rm g}\right)^{1/2}.
    \end{equation}
    \begin{equation}
        \frac{f_\text{peak}^\text{merger}}{\rm Hz}\simeq 3.5\times 10^{13}\frac{1}{B(l)A(l)^2C(l)^{3/2}}\left(\frac{M_i}{\rm g}\right)^{1/2}.
    \end{equation}
    We see that the peak frequencies scale as $\propto M_i^{1/2}$ if BH evaporated during the RD epoch.
    \item If the BH evaporated in a matter dominated (MD) Universe when $a_\text{eq}< a_\text{eva}< a_\text{de}$, with $a_\text{de}=0.8$ is the scale factor at the onset of dark energy (DE) domination, we only keep the most dominant term associated with $\Omega_m$ to get
    \begin{equation}
        a_\text{eva}\simeq 2.03\times 10^{-30}\frac{1}{B(l)^{4/3}A(l)^{8/3}}\left(\frac{M_i}{\rm g}\right)^2.
    \end{equation}
    This gives the peak frequencies:
    \begin{equation}\label{eq:fpeak_graviton_MD}
        \frac{f_\text{peak}^\text{graviton}}{\rm Hz}\simeq 2.03\times 10^7\frac{1}{B(l)^{4/3}A(l)^{5/3}}\left(\frac{M_i}{\rm g}\right).
    \end{equation}
    \begin{equation}
        \frac{f_\text{peak}^{\text{merger}}}{\rm Hz}\simeq 8.93\times 10^6\frac{1}{B(l)^{4/3}A(l)^{8/3}C(l)^{3/2}}\left(\frac{M_i}{\rm g}\right).
    \end{equation}
    We see that the peak frequencies scale as $\propto M_i$ if BH evaporated during the MD epoch.
    \item If the BH evaporated in a DE dominated Universe or is still evaporating today with $a_\text{eva}>a_\text{de}$, the redshift effect is negligible since $a_\text{de}\approx 0.8$, which corresponds to redshift $z\approx 0.3$ or 10 billion years after the Big Bang. In this case, we can simply use the formulae of frequencies in the source frame discussed earlier.
\end{itemize}

To summarize, the observed peak frequencies depend on both the genuine BH physics in the source frame and the redshift effect of cosmological background. The peak frequencies scale as $\propto M_i^{-1}$ in the source frame but scale as $\propto M_i^{1/2}$ ($\propto M_i$) if BHs evaporated in the RD (MD) phase respectively. Physically, more mass means longer lifetime and thus BHs decay later in the cosmological history, so the frequency is redshifted less. Because a MD Universe expands faster ($a(t)\propto t^{2/3}$) than a RD Universe ($a(t)\propto t^{1/2}$), the frequency is redshifted more and hence the scaling $\propto M_i$ versus $\propto M_i^{1/2}$ (see Fig. \ref{fig:fpeak_Mi}).

This analytical treatment also gave us useful insights into the relative contributions of BH's physical factors to the observed frequency. Contribution from temperature is characterized $A(l)$, from horizon size is characterized by $B(l)$, from ISCO radius is characterized by $C(l)$. The case of singular Schwarzschild BH is simply recovered in the limit $A(l\rightarrow0)=B(l\rightarrow0)=C(l\rightarrow0)=1$.

\subsection{Where to look?}\label{subsec:where to look?}

A natural question is which frequency band can we have possible maximum GW signals from both PBHs' mergers and evaporation to observe the correlation. The most promising case is PBH evaporated during RD phase before Big Bang Nucleosynthesis (BBN), preferably with PBH domination so that the maximum signal $\Omega_\text{GW,peak}^\text{graviton}\sim 10^{-7}$ could be reached (see Appendix \ref{Appendix: graviton}), and high density of PBH is also optimal for strong GW from PBH mergers \cite{Inomata:2020lmk,Hooper:2019gtx,Hooper:2020evu}\footnote{The merged PBHs will have extended lifetime, but that will not extend the PBH domination period if the merged fraction is less than $35\%$ \cite{Inomata:2020lmk}. This could be translated into an upper bound on the initial PBH fraction which is actually above the $\Delta N_\text{eff}$ constraint from isocurvature GW \cite{Papanikolaou:2020qtd,Domenech:2020ssp}, so a period of PBH domination satisfying the usual isocurvature constraint automatically satisfies this condition. For that same reason, mergers of PBH cannot finish well before evaporation and thus there is no extra redshift on $f_\text{peak}^\text{merger}$ \cite{Inomata:2020lmk}.}. Thus, we need to look for $M_i\sim 10^{-5}\rm g-10^9\rm g$ which corresponds to $f_\text{peak}^\text{graviton}\sim 10^{11}\ \rm Hz-10^{18}\ \rm Hz$ (see Eq. \ref{eq:fpeak_graviton_RD}). Recall that only PBHs with mass below $\sim 10^9\rm g$ can have PBH domination since they reheat to RD phase before BBN \cite{Loc:2025mzc}. Assuming PBHs formed from inflationary perturbation, the lower bound on frequency could be pushed up to $\sim 10^{14}\ \rm Hz$ as $M_i>\mathcal{O}(1)\rm g$ due to upper bound of tensor-to-scalar ratio \cite{Loc:2025mzc}\footnote{Other formation channels of PBHs are not subject to this bound. However, we will specifically choose the well-motivated inflationary formation channel to narrow down the frequency target.}. It will also be interesting to extend the upper frequency limit to $\sim 10^{20}\ \rm Hz$ as we may see a characteristic ``premature" GW spectrum of $10^{17}\rm g$ PBHs (see Eq. \ref{eq:fpeak_graviton_source frame}), which could be a DM candidate, with $\Omega_\text{GW,peak}^\text{graviton}\sim 10^{-10}$ (see Appendix \ref{Appendix: graviton}).

To summarize, $10^{14}\ \rm Hz-10^{20}\ \rm Hz$ is the golden frequency ballpark. PBHs evaporated during RD phase but after BBN or during MD phase give even higher frequency (see Fig. \ref{fig:fpeak_Mi}) and their population is also strongly constrained by other evaporation-related constraints \cite{Carr:2020gox} and thus the amplitude of GW spectrum is very small anyway, so any frequency higher than $10^{20}\ \rm Hz$ is not worth fighting for. At the other end, frequency below $10^{14}\ \rm Hz$ corresponds to either PBH mass below 1 gram that violates inflationary constraint or very heavy BHs that produce vanishing GW from gravitons.

\section{Application to an illustrative model: the Hayward metric}\label{sec:example}
After laying out the general framework, let us now work out an explicit example for demonstration. We consider the self-similar version \cite{Loc:2025mzc} of the Hayward metric with the metric functions given by \cite{Hayward:2005gi}
\begin{equation}
\begin{aligned}
    f_\text{Hayward}(r,l)&=g_\text{Hayward}(r,l)=1-\frac{2(r/GM)^2}{(r/GM)^3+2l^2},\\
    h_\text{Hayward}(r,l)&=r^2,
\end{aligned}
\end{equation}
where $M$ is the BH mass and $l$ is the dimensionless regularizing parameter. For $0<l<\sqrt{16/27}$, this gives a regular BH solution. The singular Schwarzschild metric is recovered when $l=0$ which gives $A(l=0)=B(l=0)=C(l=0)=1$. If $l\geq \sqrt{16/27}\approx 0.77$, the metric does not describe a BH solution. The Hayward metric features a de Sitter core that resists the total collapse of matter and is one of the earliest RBH models. It could emerge in theories of nonlinear electrodynamics \cite{Kumar:2020bqf} or from corrections due to quantum gravity effects \cite{Addazi:2021xuf}.

\begin{figure}[h!]
    \centering
    \includegraphics[width=0.45\linewidth]{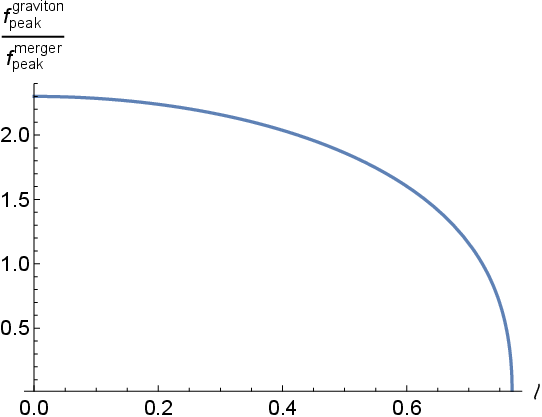}
        \includegraphics[width=0.45\linewidth]{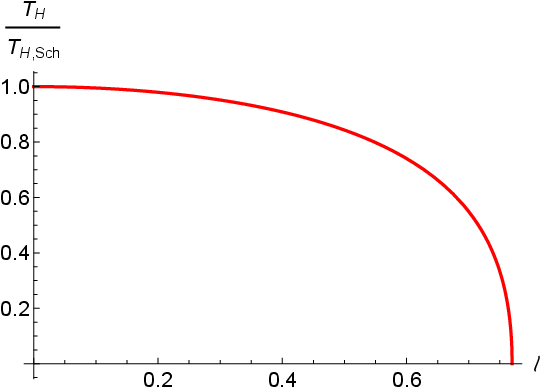}
    \includegraphics[width=0.45\linewidth]{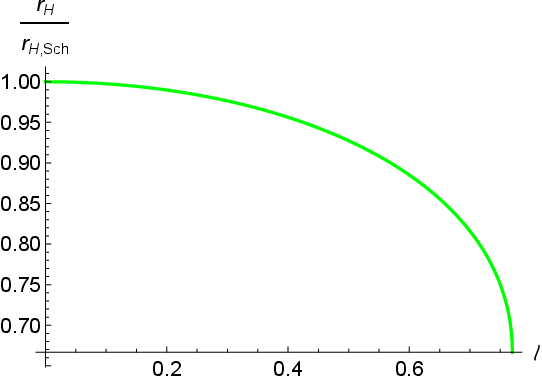}
    \includegraphics[width=0.45\linewidth]{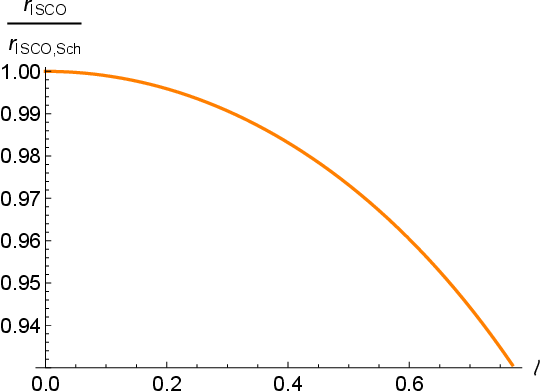}
    \caption{The observed ratio $f_\text{peak}^\text{graviton}/f_\text{peak}^\text{merger}$ can be used to infer the hair parameter $l$, along with BH's properties such as Hawking temperature, horizon size, and ISCO radius. The singular Schwarzschild BH case corresponds to $l=0$. }
    \label{fig:correlation}
\end{figure}

By using Eq. \ref{eq:frequency correlation}, we show in the upper left panel of Fig. \ref{fig:correlation} the correlation between the peak frequency of GW from gravitons $f_\text{peak}^\text{graviton}$ and the peak frequency from BH mergers $f_\text{peak}^\text{merger}$ as a function of dimensionless regularizing parameter $l$. If we observe any deviation from the expected value of $f_\text{peak}^\text{graviton}/f_\text{peak}^\text{merger}\approx 2.3$ at $l=0$, that will indicate a novel BH metric. The observed ratio can then be used to infer the hair parameter $l$ of the assumed metric, along with other BH's property such as horizon size, Hawking temperature, and ISCO radius as shown in the other panels of Fig. \ref{fig:correlation}.

\section{Comments on peak frequency of other GW sources}\label{sec: comments}

Let us now comment on some aspects of peak frequencies of other GW sources associated with PBH. The main advantage of GWs from the following sources is that the spectrum is generally at lower frequency that could be more easily accessible experimentally, but the main disadvantage is that the peak frequencies either rely on the specifics of PBH formation mechanism or are currently incapable of going beyond the linear regime of perturbation theory and thus are not conclusive for the purpose of probing genuine BH physics.

\subsection{GWs from PBH formation}

The formation of PBH is generally triggered by enhanced cosmological perturbation at small scales. Typically, these enhanced perturbations could be due to a temporary ultra slow-roll phase of the inflaton potential which serves to amplify a particular perturbation mode than average, or due to bubble collisions or collapse of Fermi balls in a first-order phase transition (FOPT) which crush the matter to the limit of BH formation. Either mechanism is violent enough to induce detectable GW signals. In the case of inflation, due to the nonlinear nature of Einstein field equation, the scalar mode is coupled with the tensor mode at second-order in perturbation theory. Therefore, the perturbation mode associated with PBH formation would induce the so-called scalar-induced GW (SIGW) upon horizon reentry \cite{Kinney:2005vj,Biagetti:2021eep,Carr:1975qj,Kohri:2018qtx,Kohri:2018awv,Domenech:2021ztg,Loc:2024qbz}. In the case of FOPT,  perturbations from bubble collisions, sound waves, or turbulence can also produce GWs \cite{Caprini:2015zlo}.

The peak frequency of GW associated with FOPT depends on \textit{both} the PBH mass and population (e.g., see Eq. 40 of \cite{Loc:2024qbz}). This peak frequency is therefore not a clear indicator of BH physics as it will also depend on the specifics of FOPT's parameters to determine the exact population for a fixed BH mass. So this candidate is not ideal to capture pure BH physics.

The SIGW spectrum peaks at the wave number corresponding to PBH formation and is given by (e.g., see Eq. 16 of \cite{Loc:2024qbz}) $
    \frac{k_\text{peak}^\text{SIGW}}{\rm Mpc^{-1}}\simeq 1.63 \times 10^{23}\gamma^{1/2}\left(\frac{\rm g}{M_i}\right)^{1/2},
$ with $\gamma\simeq 0.2$ for the standard Schwarzschild BH, or in the frequency language as $
    \frac{f_\text{peak}^\text{SIGW}}{\rm Hz}\simeq 2.53\times 10^8\gamma^{1/2}  \left(\frac{\rm g}{M_i}\right)^{1/2}.
$
The nice thing about this peak frequency is that it is much lower compared to GW from gravitons or GW from mergers, which makes it easier to detect. In addition, it is also independent from PBH population and thus does not rely on the specifics of the ultra slow-roll phase. However, the price is that we must assume both the known and the new BH solutions have the same initial condition and share the same formation mechanism. Furthermore, if there was PBH domination for ultra-light PBHs, which is an ideal situation to produce strong correlated GWs from gravitons or mergers, the SIGW signal could be suppressed by some orders of magnitude and its peak frequency will also be shifted toward lower frequency with the amount depending on the duration of PBH domination. The duration of PBH domination in turn depends on the initial PBH fraction which carries extra cosmological information that contaminates the real features of BH physics. Such issues are not present for the cases of GW from gravitons or mergers as the peak frequencies genuinely capture the BH physics, whereas the amplitude of GW is not suppressed even if there was PBH domination as they are mainly produced near the end of PBHs' evaporation and do not experience the previous PBH dominated phase.

\subsection{GWs from isocurvature perturbation}

If there was PBH domination, the rapid decay of PBH would convert the initial isocurvature perturbation between PBH and radiation into curvature perturbation that induces a second-order GW background \cite{Papanikolaou:2020qtd,Domenech:2020ssp,Loc:2025whx}. The peak frequency associated with this GW background scales as $f_\text{UV}\propto M_i^{-5/6}$ independent from PBH population. However, this frequency cutoff is not really a physical cutoff as it originates from the fact that we currently don't have available machinery to probe the scales smaller than the mean distance between nearby PBHs. The formalism also suffers an issue that $\delta\rho/\rho$ of PBH grows during PBH domination and may exceed unity before PBH's evaporation. Therefore, the linear perturbation framework used to derive the GW spectrum is no longer reliable. This kind of GW is mostly useful to obtain conservative upper bound on the population of ultra-light PBHs from $\Delta N_\text{eff}$ constraint, but the detailed GW spectrum itself does not tell conclusive physical statements.

\subsection{GWs from sudden transition}

Another source of GW from the so-called ``poltergeist mechanism" may arise if there was PBH domination, or more generally any period of early matter domination (EMD) that rapidly decays into radiation before the onset of BBN \cite{Inomata:2019ivs,Inomata:2025wiv,Inomata:2020lmk}. The basic idea is that the 
sudden transition of the gravitational potential from constancy during EMD to rapid oscillations during subsequent RD phase would induce appreciable GWs. This mechanism in general does not need the assumption of enhanced adiabatic perturbations at small scales, though we may still need it anyway if PBH forms from inflationary perturbations (at smaller scale). The peak frequency of this GW source is determined from $f_\text{peak}=\min\{f_\text{dom},f_\text{NL}\}$ \cite{Inomata:2019ivs,Allahverdi:2025btx}, where $f_\text{dom}$ corresponds to the mode entering the horizon at the onset of EMD and $f_\text{NL}$ is the mode that would become nonlinear at the end of EMD. Similar to the case of GW from isocurvature perturbation mentioned earlier, this framework only gives a conservative estimate on the strength of GW background as it cannot go beyond the linear regime of perturbation theory. If PBH domination does not last long, the GW spectrum is reliable and $f_\text{dom}$ is the physical cutoff. But then $f_\text{dom}$ depends on the onset of PBH domination, which in turn depends on the initial PBH fraction $\beta$ and thus contains some extra uncertainty that is not relevant for pure BH physics. On the other hand, if PBH domination lasts long enough such that $\delta\rho/\rho$ exceeds unity, the spectrum is cut off at $f_\text{NL}$ as modes entering horizon earlier would become nonlinear by the end of EMD and therefore cannot be described by the linear perturbation framework that was used in the first place. Thus, we conclude that the peak frequency of this GW source cannot provide clear information of BH physics.

\section{Conclusion}\label{sec:conclusion}

We showed that the correlation between peak frequencies of GW spectra from BHs' evaporation and mergers can be used to search for new BH physics. In particular, we expect $f_\text{peak}^\text{graviton}/f_\text{peak}^\text{merger}\simeq 2.3$ for the standard Schwarzschild BHs. Deviation from this ratio would indicate a new BH solution. The sweetest frequency ballpark to look for this correlation is $10^{14}\ \rm Hz-10^{20}\ \rm Hz$ as this is the optimal range to possibly have strong GW signals from both BHs' evaporation and mergers. We then applied the idea to an example of regular Hayward BH and the given general framework can be applied straightforwardly to any other self-similar type of BH.

We also commented on the potential of correlating peak frequencies among other PBH-related GW sources such as GWs associated with PBH formation, isocurvature GWs induced by PBHs' decay, and GWs induced by sudden change of the gravitational potential. The last two mechanisms are only possible if there is PBH domination. A key takeaway is that these GW sources generally peak at lower frequencies that could be more easily accessible experimentally, but the peak frequencies themselves either rely on the specifics of the formation mechanism or are unreliable for prolonged PBH domination period that may enter the nonlinear regime of perturbation theory. We thus conclude that these peak frequencies do not have sufficient predictive power to eliminate cosmological factors to capture genuine BH physics.

\section*{Acknowledgements}

This work was supported by the National Science and Technology Council, the Ministry of
Education (Higher Education Sprout Project NTU-114L104022-1), and the National Center
for Theoretical Sciences of Taiwan.

\appendix

\section{Innermost stable circular orbit}\label{appendix: rISCO}

Here we review the standard way to obtain the innermost stable circular orbit (ISCO) using the effective potential method (e.g., see \cite{Carroll:2004st}). Consider a general static, spherically symmetric spacetime of the form
\begin{equation}
    ds^2=-f(r,l)dt^2+\frac{dr^2}{g(r,l)}+h(r,l)d\Omega^2,
\end{equation}
where $d\Omega^2=d\theta^2+(\sin\theta)^2d\phi^2$ is the usual 2-sphere metric. Here, $l$ is a parameter, not a variable. Invariance under time translation (static) leads to 1 Killing vector associated with energy conservation. Invariance under spatial rotation (spherical symmetry) leads to 3 Killing vectors associated with angular momentum conservation. Angular momentum conservation could be separated into conservation of magnitude and direction, the latter of which implies that the test particle moves on a fixed plane. Because of spherical symmetry, we can always choose the coordinates such that the particle moves on the equatorial plane with $\theta=\pi/2$. The remaining two Killing vectors are 
\begin{equation}
    K^\mu=(\partial_t)^\mu=\delta^\mu_t=(1,0,0,0)\Rightarrow K_\mu=(-f(r,l),0,0,0)
\end{equation}
\begin{equation}
    R^\mu=(\partial_\phi)^\mu=\delta_\phi^\mu=(0,0,0,1)\Rightarrow R_\mu=(0,0,0,h(r,l))
\end{equation}
The conserved quantities are $K_\mu dx^\mu/d\tau$, which give\footnote{For massive particle, the affine parameter could naturally be chosen to be the proper time $\tau$.}
\begin{equation}
    E=-K_\mu\frac{dx^\mu}{d\tau}=f(r,l)\frac{dt}{d\tau}
\end{equation}
\begin{equation}
    L=R_\mu\frac{dx^\mu}{d\tau}=h(r,l)\frac{d\phi}{d\tau}
\end{equation}
Here, $E$ and $L$ are conserved energy and angular momentum per unit mass respectively. By using the condition $U^\mu U_\mu=-1$ for massive particle with $U^\mu=dx^\mu/d\tau$ is the 4-velocity, we get
\begin{equation}
    \frac{f(r,l)}{g(r,l)}\left(\frac{dr}{d\tau}\right)^2=E^2-V_\text{eff}(r,l),
\end{equation}
where
\begin{equation}
    V_\text{eff}(r,l)\equiv f(r,l)\left[\frac{L^2}{h(r,l)}+1 \right].
\end{equation}
To find ISCO, we set $dr/d\tau=0$ corresponding to fixed radial distance. This implies the first condition $V_\text{eff}(r_\text{ISCO},l)=E^2$, or
\begin{equation}
    V_\text{eff}'(r,l)\bigg|_{r=r_\text{ISCO}}=0
\end{equation}
\begin{equation}
    \Rightarrow L^2=\frac{f'h^2}{fh'-f'h}
\end{equation}
where prime denotes derivative with respect to $r$ and we temporarily omitted the appearance of $r$ and $l$ for brevity. In general relativity, ISCO corresponds to a local maximum of the effective potential, so the second condition is
\begin{equation}
    V_\text{eff}''(r,l)\Bigg|_{r=r_\text{ISCO}}=0
\end{equation}
which marks the transition from stability to instability. Combining with the first condition, this gives 
\begin{equation}\label{eq:ISCO master}
    f''fh'-2f'^2h'+f'f\left(\frac{2h'^2}{h}-h''\right)=0.
\end{equation}
This is the final general result that we can solve for $r_\text{ISCO}$ as a function of $l$ (if $l\neq 0$).

For example, the standard Schwarzschild metric has
\begin{equation}
    f_\text{Sch}=1-\frac{2GM}{r},\ h_\text{Sch}=r^2,
\end{equation}
which give the well-known result
\begin{equation}
    r_\text{ISCO,Sch}=6GM
\end{equation}
For more complicated metrics, including the Hayward metric used in this paper, Eq. \ref{eq:ISCO master} may have to be solved numerically. It has been shown that $r_\text{ISCO}\leq 6GM$ for a general static, spherically symmetric spacetimes with matter fields satisfying the weak energy condition \cite{Cen:2025atm}. Indeed, $r_\text{ISCO}$ of the Hayward metric satisfies this upper bound (see Fig. \ref{fig:correlation}).

\section{GW from gravitons}\label{Appendix: graviton}

To determine the peak frequency of GW spectrum made of gravitons, we first need the functional form of the spectrum. The fractional energy density of GW today is defined as
\begin{equation}
    \Omega_\text{GW,0}(f)\equiv\frac{1}{\rho_\text{crit,0}}\frac{d\rho_\text{GW,0}}{d\ln f}=\frac{1}{\rho_\text{crit,0}}\frac{d\rho_\text{GW,0}}{d\ln\omega_0},
\end{equation}
where $\rho_\text{crit,0}=3H_0^2/8\pi G$ is the current critical energy density with $H_0=67\ \rm km/s/Mpc^{-1}$ is the Hubble constant \cite{Planck:2018jri}, $\rho_\text{GW,0}$ is the current energy density of GW, and $\omega_0=2\pi f$
is the current frequency of gravitons. Since $\omega\propto a^{-1}$ and $\rho_\text{GW}\propto a^{-4}$ with $a$ is the scale factor, we have
\begin{equation}
    \frac{d\rho_\text{GW,0}}{d\ln\omega_0}=\omega_0\frac{d\rho_\text{GW,0}}{d\omega_0}=\omega_\text{eva}\frac{d\rho_\text{GW,eva}}{d\omega_\text{eva}}\left(\frac{a_\text{eva}}{a_0}\right)^4.
\end{equation}
Let $n_\text{PBH}(t)$ be the instantaneous number density of PBH at time $t$ and $E_\text{grav}$ be the total energy of gravitons radiated per BH, we have
\begin{equation}
    \frac{d^2\rho_\text{GW}(t)}{dtd\omega}=n_\text{PBH}(t)\frac{d^2E_\text{grav}}{dtd\omega}.
\end{equation}
Performing the integration over time, we have
\begin{equation}
    \frac{d\rho_\text{GW,eva}}{d\omega_\text{eva}}=\int_{t_i}^{t_\text{eva}}  n_\text{PBH}(t)\frac{d^2E_\text{grav}}{dtd\omega}dt
    =\int_{t_i}^{t_\text{eva}}n_\text{PBH,i}\left(\frac{a_i}{a(t)}\right)^3\frac{d^2E_\text{grav}}{dtd\omega}dt,
\end{equation}
where $t_\text{eva}=t_i+\tau$. If $t_\text{eva}>t_\text{age}^\text{Universe}(a_0)\approx 13.8\rm \ Gyr$, the upper bound of the integral must be cut off at $t_\text{age}^\text{Universe}(a_0)$. In the blackbody approximation, we have \cite{Loc:2025mzc}
\begin{equation}
    \frac{d^2E_\text{grav}}{dtd\omega}=4\pi R_\text{H}^2\times\frac{2}{8\pi^2}\frac{\omega_\text{eva}^3}{e^{\omega/T_\text{H}}-1}=\frac{4B(l)^2M^2\omega_\text{eva}^3}{\pi m_\text{pl}^4(e^{\omega/T_\text{H}}-1)},
\end{equation}
where we used the parameterization in Eq. \ref{eq: A and B} and the Schwarzschild radius in Eq. \ref{eq: T_Sch and r_Sch}, and the fact that gravitons are bosons with 2 internal degrees of freedom. The GW spectrum is then
\begin{equation}
        \Omega_\text{GW,0}(\omega_0)=\frac{1}{\rho_\text{crit,0}}\omega_0^4n_\text{PBH,i}a_i^3\int_{t_i}^{t_\text{eva}}\frac{1}{a(t)^3}\frac{4B(l)^2M(t)^2}{\pi m_\text{pl}^4(e^{\omega(t)/T_\text{H}(t)}-1)}dt,
\end{equation}
where we used the fact that $\omega_\text{eva}^3a_\text{eva}^3=\omega_0^3$ with the scale factor today is set to unity as usual $a_0=1$.

We can inspect the functional form of $\Omega_\text{GW,0}(\omega_0)$ in order to determine the peak frequency. The dependence on $\omega_0$ takes the form
$
\frac{\omega_0^4}{e^{\omega_0/a_\text{eva}T_\text{H,i}}-1}
$, where we used the fact that BH decay is a very rapid process so that we can approximate $\omega(t)\rightarrow\omega_\text{eva}$ and $T_\text{H}(t)\rightarrow T_\text{H,i}$ \footnote{This approximation is only used to analytically determine the peak frequency. The actual spectrum uses the exact time-dependent quantities and is calculated numerically.}
. By extremizing  this function with respect to $\omega_0$, this leads to the equation of the form
$
\left(1-\frac{x}{4}\right)e^x-1=0
$
with $x\equiv \omega_0/a_\text{eva}T_\text{H,i}$ which admits only one nontrivial solution $x=3.92$. If the sources are nearby, we have $a_\text{eva}=a_0=1$ and use Eq. \ref{eq:T_H} to get \footnote{If PBHs have high spin, the peak frequency will be slightly shifted towards lower frequency as the rotating BH has lower Hawking temperature. However, PBHs with negligible spin are expected if formed during the standard RD Universe.}
\begin{equation}
    \left(\frac{\omega_0^\text{peak}}{\rm GeV}\right)=3.92\left(\frac{T_\text{H,i}}{\rm GeV}\right)\simeq 4.16\times 10^{13}A(l)\left(\frac{\rm g}{M_i}\right).
\end{equation}
By using the conversion relation $\omega_0=2\pi f$, or
\begin{equation}
    \left(\frac{\omega_0}{\rm GeV}\right)=4.15\times 10^{-24}\left(\frac{f}{\rm Hz}\right),
\end{equation}
the peak frequency in the conventional Hertz language is
\begin{equation}\label{eq:f graviton peak}
    \left(\frac{f_\text{peak}^\text{graviton}}{\rm Hz}\right)\simeq 10^{37}A(l)\left(\frac{\rm g}{M_i}\right),
\end{equation}
which is Eq. \ref{eq:fpeak_graviton_source frame} in the main text \footnote{The peak frequency is not affected by the inclusion of greybody factors as it is mainly determined by the Hawking temperature \cite{Ireland:2023avg}.}.

While further calculating the GW spectrum is not necessary for the present purpose of this paper, we think it would be useful to record these results here to gain some interesting physical insights. Recall that the PBH fraction at the formation time in a RD phase is defined as $\beta\equiv\rho_\text{PBH,i}/\rho_\text{rad,i}$, which characterizes population of PBH for a given mass. Since PBHs' energy density is diluted slower than radiation, $\rho_\text{PBH}\propto a^{-3}$ versus $\rho_\text{rad}\propto a^{-4}$, there exists a critical value of PBH fraction at the formation time above which there will be PBH domination \cite{Loc:2025mzc}
\begin{equation}
    \beta_c\simeq 1.78\times 10^{-6}\gamma^{-1/2}A(l)^2B(l)\left(\frac{g_*(T_\text{H})}{106.75}\right)^{1/2}\left(\frac{106.75}{g_{*,\rm eva}}\right)^{1/4}\left(\frac{g_{*,\rm i}}{106.75}\right)^{1/4}\left(\frac{\rm g}{M_i}\right).
\end{equation}
However, a period of PBH domination is only possible if it terminates before the BBN time $t_\text{eva}<t_\text{BBN}\sim 1\ \rm sec$. We have two cases:

(1) If $\beta<\beta_c$, there is no PBH domination and thus entropy is conserved. We then note that
    \begin{equation}
        n_\text{PBH,i}a_i^3=\frac{\beta}{M_i}\rho_\text{rad,i}a_i^3=\frac{\beta}{M_i}\frac{\pi^2}{30}g_{*,i}T_i^4a_i^3=\frac{\beta}{M_i}\frac{\pi^2}{30}T_ig_{s,0}T_0^3a_0^3
    \end{equation}
    The GW spectrum today is then
    \begin{equation}
            \begin{aligned}
        \Omega_\text{GW,0}
        &=\frac{8\pi}{3H_0^2}\frac{\omega_0^4}{m_\text{pl}^2}\frac{\beta}{M_i}\frac{\pi^2}{30}T_ig_{s,0}T_0^3a_0^3\int_{t_i}^{t_\text{eva}}\frac{1}{a(t)^3}\frac{4B(l)^2M_i^2\left(1-\frac{t-t_i}{\tau}\right)^{2/3}}{\pi m_\text{pl}^4\left(\exp \left[\frac{\omega_0}{a(t)}\frac{8\pi}{A(l)m_\text{pl}^2}M_i\left(1-\frac{t-t_i}{\tau} \right)^{1/3} \right] -1\right)}dt\\
        &=\frac{16\pi^2}{45}\frac{\omega_0^4}{H_0^2m_\text{pl}^6}\beta M_i\left(\frac{\gamma\sqrt{45}m_\text{pl}^3}{4\pi^{3/2}g_{*,i}^{1/2}M_i}\right)^{1/2}g_{s,0}T_0^3B(l)^2 I(\omega_0)\\
        &=\frac{8\pi^{5/4}}{45^{3/4}}\frac{\omega_0^4}{H_0^2m_\text{pl}^{9/2}}\beta M_i^{1/2}\frac{\gamma^{1/2}}{g_{*,i}^{1/4}}g_{s,0}T_0^3B(l)^2I(\omega_0)\\
        &=2.06\times 10^{-4}\gamma^{1/2}B(l)^2\left(\frac{106.75}{g_{*,i}}\right)^{1/4}\left(\frac{M_i}{\rm g}\right)^{1/2}\beta \left(\frac{\omega_0}{\rm GeV}\right)^4\left(\frac{I(\omega_0)}{\rm sec}\right),\label{eq:beta<beta_c}
    \end{aligned}
        \end{equation}
    where 
    \begin{equation}
        I(\omega_0)\equiv \int_{t_i}^{t_\text{eva}}\frac{1}{a(t)^3}\frac{\left(1-\frac{t-t_i}{\tau}\right)^{2/3}}{\exp \left[\frac{\omega_0}{a(t)}\frac{8\pi}{A(l)m_\text{pl}^2}M_i\left(1-\frac{t-t_i}{\tau}\right)^{1/3} \right]-1}dt
    \end{equation}
    \begin{equation}
    \Rightarrow
    \frac{I(\omega_0)}{\rm sec}=\int_{t_i}^{t_\text{eva}}\frac{1}{a(t)^3}\frac{(1-\frac{t-t_i}{\tau})^{2/3}}{\exp\left[9.45\times 10^{-14}\frac{1}{a(t)}\frac{1}{A(l)}\left(\frac{\omega_0}{\rm GeV}\right)\left(\frac{M_i}{\rm g}\right)\left(1-\frac{t-t_i}{\tau}\right)^{1/3} \right]-1}dt.
\end{equation}
As there is no PBH domination in this case, we can use the usual scale factor of standard cosmology. We employ the piecewise analytical formulae that accurately matches the exact numerical solution given in \cite{Galanti:2021bxt}.

(2) If $\beta>\beta_c$, there is PBH domination and entropy is only conserved after PBH evaporation. By noting that $n_\text{PBH}\propto a^{-3}$, $\rho_\text{PBH,eva}\sim\rho_\text{rad,eva}=\pi^2g_{*,\text{eva}}T_\text{eva}^4/30$ and entropy is conserved from evaporation time to today so that $g_\text{s,eva}T_\text{eva}^3=g_\text{s,0}T_0^3$, we have
    \begin{equation}
    n_\text{PBH,i}a_i^3=n_\text{PBH,eva}a_\text{eva}^3=\frac{\rho_\text{rad,eva}a_\text{eva}^3}{M_i}=\frac{\pi^2}{30}g_{*,\rm eva}T_\text{eva}^4\frac{a_\text{eva}^3}{M_i}=\frac{\pi^2}{30}g_{*,\rm eva}\left(\frac{g_{s,0}}{g_{s,\rm eva}}\right)^{4/3}\frac{T_0^4}{a_\text{eva}^4}\frac{a_\text{eva}^3}{M_i}
    \end{equation}
    The GW spectrum today is then
    \begin{equation}
    \begin{aligned}
        \Omega_\text{GW,0}&=\frac{8\pi}{3H_0^2}\frac{\omega_0^4}{m_\text{pl}^2}\frac{\pi^2}{30}g_{*,\rm eva}\left(\frac{g_{s,0}}{g_{s,\rm eva}}\right)^{4/3}\frac{T_0^4}{a_\text{eva}^4}\frac{a_\text{eva}^3}{M_i}\times\\
        &\times\int_{t_i}^{t_\text{eva}}\frac{1}{a(t)^3}\frac{4B(l)^2M_i^2\left(1-\frac{t-t_i}{\tau}\right)^{2/3}}{\pi m_\text{pl}^4\left( \exp \left[\frac{\omega_0}{a(t)}\frac{8\pi}{A(l)m_\text{pl}^2}M_i\left(1-\frac{t-t_i}{\tau} \right)^{1/3} \right]-1\right)}dt\\
        &=\frac{16\pi^2}{45}\frac{\omega_0^4}{H_0^2m_\text{pl}^6}\frac{g_{s,0}^{4/3}}{g_{*,\rm eva}^{1/3}}T_0^4\frac{M_i}{a_\text{eva}}B(l)^2I(\omega_0)\\
        &=1.62\times 10^{-33}\frac{B(l)^2}{a_\text{eva}}\left(\frac{106.75}{g_{*,\rm eva}}\right)^{1/3}\left(\frac{M_i}{\rm g}\right)\left(\frac{\omega_0}{\rm GeV}\right)^4\left(\frac{I(\omega_0)}{\rm sec}\right)\\
        &=3.66\times 10^{-10}A(l)^2B(l)^3\left(\frac{106.75}{g_{*,\rm eva}}\right)^{1/4}\left(\frac{g_*(T_H)}{106.75}\right)^{1/2} \left(\frac{\rm g}{M_i}\right)^{1/2}\left(\frac{\omega_0}{\rm GeV}\right)^4\left(\frac{I(\omega_0)}{\rm sec}\right),\label{eq:beta>beta_c}
    \end{aligned}
    \end{equation}
    where $I(\omega_0)$ is defined as before and Eq. \ref{eq:a_eva} for $a_\text{eva}$ was used. We see that the GW amplitude no longer depends on PBH population characterized by $\beta$. The origin of this independence comes from the fact that we gauged $\rho_\text{PBH}\sim\rho_\text{rad}$ at the evaporation time which is only possible if there is PBH domination. The scale factor in this case has two evolutionary stages, $a\propto t^{1/2}$ before PBH domination and $a\propto t^{2/3}$ during PBH domination:
    \begin{equation}
        a(t)=\begin{cases}
            a_\text{dom}\left(\frac{t}{t_\text{dom}}\right)^{1/2}\hspace{1cm}t_i\leq t\leq t_\text{dom} \\
            a_\text{eva}\left(\frac{t}{\tau}\right)^{2/3}\hspace{1cm} t_\text{dom}\leq t\leq t_\text{eva}
        \end{cases},
    \end{equation}
    where the subscript ``dom" indicates the quantities at the onset of PBH domination. Because $\rho_\text{PBH}=\rho_\text{rad}$ at $a_\text{dom}$, we have $\beta=a_i/a_\text{dom}=(t_i/t_\text{dom})^{1/2}$. By using Eq. \ref{eq:ti}, we then have
    \begin{equation}
        \left(\frac{t_\text{dom}}{\rm sec}\right)=\beta^{-2}\left(\frac{t_i}{\rm sec}\right)=2.48\times 10^{-39}\gamma^{-1}\left(\frac{M_i}{\rm g}\right)\beta^{-2}.
    \end{equation}
    The matching condition at $t_\text{dom}$ gives $a_\text{dom}=a_\text{eva}(t_\text{dom}/\tau)^{2/3}$. By using Eq. \ref{eq:a_eva}, we then have
    \begin{equation}
        a_\text{dom}=6\times 10^{-32}\gamma^{-2/3}A^{2/3}B^{1/3}\left(\frac{106.75}{g_{*,\rm eva}}\right)^{1/12}\left(\frac{g_*(T_H)}{106.75}\right)^{1/6} \left(\frac{M_i}{\rm g}\right)^{1/6}\beta^{-4/3}.
    \end{equation}
    Thus, the final expression for us to use is
    \begin{equation}
        a(t)=\begin{cases}
            1.21\times 10^{-12}\gamma^{-1/6}A^{2/3}B^{1/3}\left(\frac{106.75}{g_{*,\rm eva}}\right)^{1/12}\left(\frac{g_*(T_H)}{106.75}\right)^{1/6} \left(\frac{\rm g}{M_i}\right)^{1/3}\beta^{-1/3}\left(\frac{t}{\rm sec}\right)^{1/2}\hspace{1cm}t_i\leq t\leq t_\text{dom} \\
            3.28\times 10^{-6}A^{2/3}B^{1/3}\left(\frac{106.75}{g_{*,\rm eva}}\right)^{1/12}\left(\frac{g_*(T_H)}{106.75}\right)^{1/6} \left(\frac{\rm g}{M_i}\right)^{1/2}\left(\frac{t}{\rm sec}\right)^{2/3}\hspace{1cm}t_\text{dom}\leq t\leq t_\text{eva}
        \end{cases}
    \end{equation}

\begin{figure}[h!]
    \centering
    \includegraphics[width=0.49\linewidth]{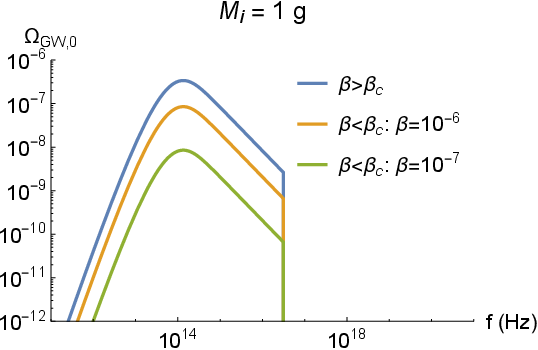}
    \includegraphics[width=0.49\linewidth]{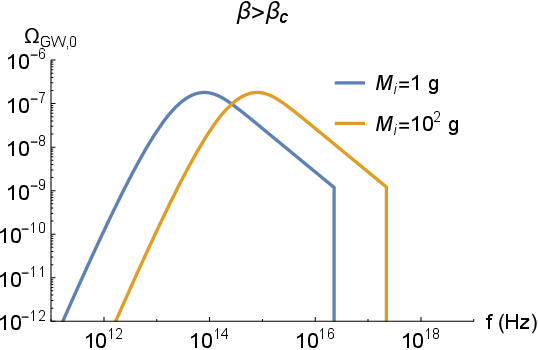}\vspace{0.5cm}
    \includegraphics[width=0.49\linewidth]{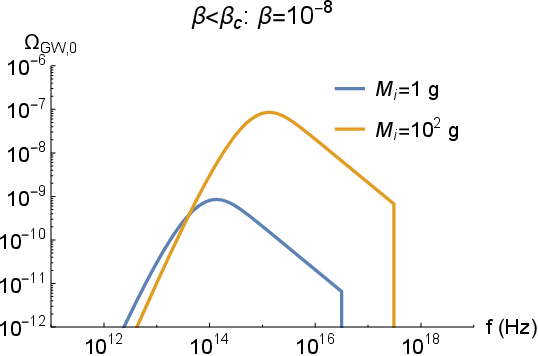}
    \includegraphics[width=0.49\linewidth]{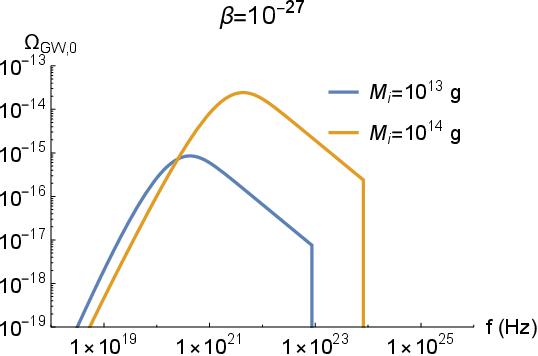}\vspace{0.5cm}
    \includegraphics[width=0.49\linewidth]{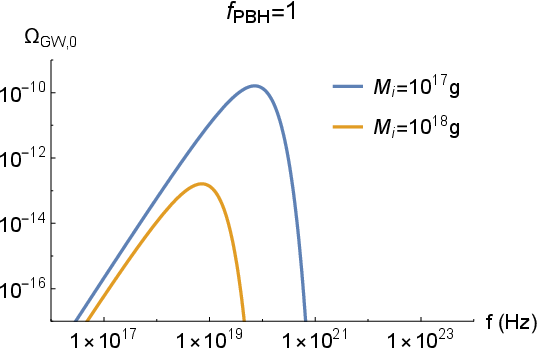}
    \includegraphics[width=0.49\linewidth]{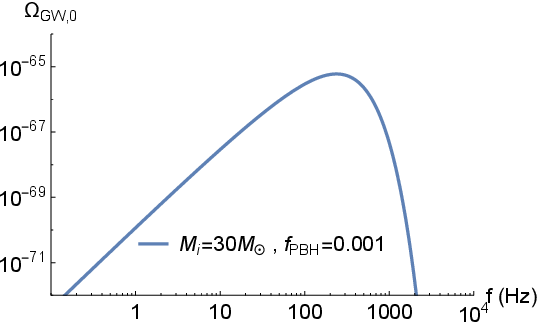}
    \caption{GW spectrum of gravitons produced from evaporation of Schwarzschild PBH. See texts for descriptions and physical interpretations.}
    \label{fig:GW spectrum}
\end{figure}

We show in Fig. \ref{fig:GW spectrum} the GW spectrum of gravitons produced from evaporation of Schwarzschild PBH corresponding to $A(l=0)=B(l=0)=1$ and $\gamma=0.2$. The first general remark is that the form of GW spectrum of fully evaporated PBHs changes from $\Omega_\text{GW}\propto f^{5/2}$ for $f<f_\text{peak}$ to $\Omega_\text{GW}\propto f^{-1}$ for $f>f_\text{peak}$ because a large number of high-energy gravitons is produced at the late stage of evaporation when PBH gets smaller and hotter. The vertical drop of the spectrum corresponds to the moment when PBHs have been evaporated completely. This is a very distinctive form of GW spectrum made of gravitons that could be discriminated from other GW sources \footnote{Note that while the form of GW spectrum associated with PBH formation is dictated by the physics of cosmological perturbation that made PBH which is model-dependent, the form of GW spectrum made of gravitons is model-independent as it only involves the physics of BH's evaporation. Also, our approximation captures accurately the spectral form calculated using the full greybody factors with very little difference in amplitude (e.g., see \cite{Choi:2024acs}). In \cite{Ireland:2023avg}, the authors got the wrong spectral shape and wrong amplitude when using the blackbody approximation because they mistakenly took the time-dependent mass term $M(t)^2$ in the numerator outside the time integral in their Eq. (23).}. For PBHs that are still evaporating now, the GW spectrum features a ``premature" spectral shape with $\Omega_\text{GW}\propto f^{5/2}$ for $f<f_\text{peak}$ and rapid drop for $f>f_\text{peak}$. This reflects the fact that gravitons are produced most efficiently in terms of number and energy only at the late stage of evaporation as BH decay is an explosive process near the end of its lifetime. GW spectrum made of gravitons evaporated from young BH therefore has not yet developed the high-frequency tail. Some specific remarks for each panel of Fig. \ref{fig:GW spectrum} are as follows:
\begin{itemize}
    \item \textbf{Upper left: Fixed mass and varied $\beta$, PBH evaporated during RD phase} \\
    The peak frequency is solely determined by Hawking temperature gauged by the mass, so it is independent from population. The amplitude of GW spectrum is independent from population as well if there is PBH domination with $\beta>\beta_c$ (for 1 gram PBH, $\beta_c\simeq 4\times10^{-6}$), since the dependence was removed when balancing $\rho_\text{PBH}\sim\rho_\text{rad}$ at the evaporation time (explained below Eq. \ref{eq:beta>beta_c}). However, if there is no PBH domination with $\beta<\beta_c$, the amplitude \textit{does} depend on the population (see Eq. \ref{eq:beta<beta_c}). Less population produces weaker GW background.
    
    \item \textbf{Upper right: Fixed $\beta>\beta_c$ and varied mass, PBH evaporated during RD phase} \\
    The purpose is to show that, for PBHs evaporated during RD phase, the peak frequency scales as $f_\text{peak}\propto M_i^{1/2}$. The amplitude of GW spectrum at $\Omega_\text{GW,peak}\sim 2\times10^{-7}$ is independent from PBH mass, in addition to being independent from population as mentioned above. The reason is that, at the peak, the prefactor in front of $I(\omega_0)$ in Eq. \ref{eq:beta>beta_c} goes like $\propto M_i^{-1/2}\omega_{0,\rm peak}^4\propto M_i^{3/2}$ for PBHs evaporated during RD phase (see Eq. \ref{eq:fpeak_graviton_RD}), whereas it could be checked numerically that $I(f_\text{peak})\propto M_i^{-3/2}$ and hence the cancellation. Physically, because $\rho_\text{PBH}/\rho_\text{tot}$ is constant during PBH domination, where $\rho_\text{tot}$ is the total energy density, the produced ratio $\rho_\text{GW}/\rho_\text{tot}$ is the same for different PBH masses. The ratio $\rho_\text{GW}/\rho_\text{tot}$ is frozen during the subsequent RD phase and experiences the same amount of suppression during the MD and DE phases.
    
    \item \textbf{Middle left: Fixed $\beta<\beta_c$ and varied mass, PBH evaporated during RD phase} \\
    The value $\beta=10^{-8}$ was chosen such that there is no PBH domination for both 1 gram and 100 grams PBHs. The purpose is to show that, if there is no PBH domination, $\Omega_\text{GW,peak}$ \textit{does} depend on the PBH mass. The reason is that, at the peak, the prefactor in front of $I(\omega_0)$ in Eq. \ref{eq:beta<beta_c} goes like $\propto M_i^{1/2}\beta\omega_{0,\rm peak}^4\propto M_i^{5/2}\beta$ for PBHs evaporated during RD phase (see Eq. \ref{eq:fpeak_graviton_RD}), whereas it could be checked numerically that $I(f_\text{peak})\propto M_i^{-3/2}$ and hence $\Omega_\text{GW,peak}\propto M_i\beta$. Physically, because $\rho_\text{PBH}/\rho_\text{tot}\propto a$ when there is no PBH domination, heavier PBH lives longer and thus the produced ratio $\rho_\text{GW}/\rho_\text{tot}$ must also be larger.

    \item \textbf{Middle right: Fixed $\beta$ and varied mass, PBH evaporated during MD phase.} \\
    Note that it is not possible to have PBH domination after BBN.
    The value of $\beta$ was chosen such that it satisfies CMB-related constraints \cite{Carr:2020gox}. The purpose is to demonstrate that, if PBHs evaporated during the MD phase, the peak frequency scales as $f_\text{peak}\propto M_i$. The amplitude of GW background also depends on the mass. In particular, at the peak, the prefactor in front of $I(\omega_0)$ in Eq. \ref{eq:beta<beta_c} scales as $\propto M_i^{1/2}\beta\ \omega_{0,\rm peak}^4\propto M_i^{9/2}\beta$ for PBHs evaporated during MD phase (see Eq. \ref{eq:fpeak_graviton_MD}), whereas it could be checked numerically that $I(f_\text{peak})\propto M_i^{-3}$ and thus $\Omega_\text{GW,peak}\propto M_i^{3/2}\beta$. The physical explanation is similar to the previous case when we considered $\beta<\beta_c$ with PBHs evaporated during RD phase. Namely, the fraction of total energy density in PBH grows more for heavier (thus longer-lived) PBH during RD phase (since $\rho_\text{PBH}/\rho_\text{tot}\propto a$) before it becomes frozen during MD phase (since $\rho_\text{PBH}/\rho_\text{tot}=const.$), and thus the produced ratio $\rho_\text{GW}/\rho_\text{tot}$ must be larger. But now on top of that, PBHs evaporated later during the MD phase also experience less suppression of the ratio $\rho_\text{GW}/\rho_\text{tot}
    \propto a^{-1}$, and thus $\Omega_\text{GW,peak}$ is enhanced more than in the case of evaporation during RD phase.

    \item \textbf{Lower left: PBH as a DM candidate with $f_\text{PBH}=1$ and varied PBH mass, PBH still evaporating}\\
    Note that for non-evaporated PBHs, it is more convenient to use the fraction of PBH in DM $f_\text{PBH}\equiv \Omega_\text{PBH,0}/\Omega_\text{DM,0}$ by the relation  $\beta\simeq 6\times 10^{-26}\gamma^{-1/2}(g_{*,i}/106.75)^{1/4}(M_i/ \text{g})^{1/2} f_\text{PBH}$ \cite{Loc:2024qbz}. For PBHs that have lifetime greater than the age of the Universe, the $I(\omega_0)$ integral must be cut off at $t_\text{age}^\text{Universe}$. We see that the GW spectrum is ``premature" as the high-frequency tail has not yet established the expected scaling $\Omega_\text{GW}\propto f^{-1}$ because the PBHs have not yet entered the final stage of evaporation. The peak frequency now scales as $f_\text{peak}\propto M_i^{-1}$ since there is no redshift. The amplitude of GW spectrum for heavier PBHs is now smaller because, for a fixed available cosmic time, heavier PBHs are at the earlier stage of evaporation. Mathematically, it can be seen that the prefactor in front of $I(\omega_0)$ in Eq. \ref{eq:beta<beta_c} goes like $\propto M_i^{1/2}\beta\ \omega_{0,\rm peak}^4\propto M_i^{1/2}M_i^{1/2}f_\text{PBH}M_i^{-4} \propto M_i^{-3}f_\text{PBH}$ for PBHs that are still evaporating now (see Eq. \ref{eq:fpeak_graviton_source frame}), whereas we have checked numerically that $I(f_\text{peak})$ is independent from $M_i$ and thus $\Omega_\text{GW,peak}\propto M_i^{-3}f_\text{PBH}.$ It is interesting to see that, although no currently proposed detector is able to probe such high-frequency regime of order $\sim 10^{20}\ \rm Hz$, the strength of GW is of order $\sim 10^{-10}$ which seems encouraging compared with sensitivity level of order $\sim 10^{-12}$ of typical low-frequency detectors such as LISA. Both of the facts that there is no constraint on PBH population in the asteroid-mass DM window and freshly produced GW in the nearby Universe does not experience significant redshift lead to strong GW spectrum. The hope to test both the PBH-as-DM scenario and the quantization nature of gravity fades quickly for larger mass as GW from gravitons is suppressed due to insufficient available cosmic time, and thus the low-mass end of the PBH DM window with mass $\sim 10^{17}\rm g$ is the only promising chance we have.

    \item \textbf{Lower right: The seminal case of $\sim 30M_\odot$ BHs detected in the first ever GW event GW150914 \cite{LIGOScientific:2016aoc}, BH still evaporating.}\\
    The population $f_\text{PBH}=0.001$ was chosen to touch the upper bound coming from merger rate constraint \cite{Carr:2020gox}. The purpose is to show that, although the peak frequency is right at the sensitivity of LIGO, it is impossible to see the GW spectrum as the amplitude is very small $\Omega_\text{GW,peak}\sim 10^{-65}$.
\end{itemize}
We summarize these comments in Table \ref{Table:summary}.

\begin{table}[h!]
\centering
\begin{tabular}{|c|c|c|c|}
\hline 
\multicolumn{2}{|c|}{Mass}  & $f_\text{peak}^\text{graviton}(\rm Hz)$ & $\Omega_\text{GW,peak}^\text{graviton}$         \\ 
\hline
\multirow{2}{*}{
$
\begin{aligned}
    M_i&\lesssim 10^{13}\rm g\\
(t_\text{eva}&\lesssim t_\text{eq})    
\end{aligned}$ 
}
&
$
\begin{aligned}
    \beta&>\beta_c \\
(t_\text{eva}&<t_\text{BBN}\ \rm  only)
\end{aligned}
$     
& $\simeq 8\times 10^{13}\left(\frac{M_i}{\rm g}\right)^{1/2}$ (Eq. \ref{eq:fpeak_graviton_RD})       & $\simeq 2\times 10^{-7}$ (Eq. \ref{eq:beta>beta_c})                \\ \cline{2-4} 
                   & $\beta<\beta_c$                  & $\simeq 8\times 10^{13}\left(\frac{M_i}{\rm g}\right)^{1/2}$    (Eq. \ref{eq:fpeak_graviton_RD})     & $\simeq 8\times10^{-2}\left(\frac{M_i}{\rm g}\right)\beta$   (Eq. \ref{eq:beta<beta_c})          \\ \hline
                   \multicolumn{2}{|c|}{$
                   \begin{aligned}
                       10^{13}\rm g&\lesssim M_i\lesssim 10^{15}\rm g\\
                       (t_\text{eq}&\lesssim t_\text{eva}\lesssim t_\text{age}^\text{Universe})
                   \end{aligned}
                   $} & 
$\simeq 2\times 10^7\left(\frac{M_i}{\rm g}\right)$   (Eq. \ref{eq:fpeak_graviton_MD})     & $\simeq 2\times 10^{-8}\left(\frac{M_i}{\rm g}\right)^{3/2}\beta$  (Eq. \ref{eq:beta<beta_c})             \\ \hline
\multicolumn{2}{|c|}{$
\begin{aligned}
    M_i&\gtrsim 10^{15}\rm g\\
    (t_\text{eva}&>t_\text{age}^\text{Universe})
\end{aligned}
$}
& $\simeq 10^{37}\left(\frac{\rm g}{M_i}\right)$   (Eq. \ref{eq:fpeak_graviton_source frame})       & $\simeq 2\times 10^{41}\left(\frac{\rm g}{M_i}\right)^3f_\text{PBH}$ (Eq. \ref{eq:beta<beta_c})               \\ \hline
\end{tabular}
\caption{Summary of how the peak frequency and peak amplitude of GW spectrum made of gravitons evaporated from Schwarzschild PBH depend on PBH's mass and population.}
\label{Table:summary}
\end{table}

\bibliographystyle{JHEP}
\bibliography{references}

\providecommand{\href}[2]{#2}\begingroup\raggedright\begin{thebibliography}{10}

\bibitem{Bardeen}
J.~Bardeen{\emph{Proceedings, 5th International Conference on Gravitation and the theory of relativity: Tbilisi, USSR, December 9-13, 1968 (1968) p. 174.} }.

\bibitem{Hayward:2005gi}
S.A.~Hayward, \emph{{Formation and evaporation of regular black holes}}, \href{https://doi.org/10.1103/PhysRevLett.96.031103}{\emph{Phys. Rev. Lett.} {\bfseries 96} (2006) 031103} [\href{https://arxiv.org/abs/gr-qc/0506126}{{\ttfamily gr-qc/0506126}}].

\bibitem{Dymnikova:1992ux}
I.~Dymnikova, \emph{{Vacuum nonsingular black hole}}, \href{https://doi.org/10.1007/BF00760226}{\emph{Gen. Rel. Grav.} {\bfseries 24} (1992) 235}.

\bibitem{Culetu:2013fsa}
H.~Culetu, \emph{{On a regular modified Schwarzschild spacetime}},  \href{https://arxiv.org/abs/1305.5964}{{\ttfamily 1305.5964}}.

\bibitem{Culetu:2014lca}
H.~Culetu, \emph{{On a regular charged black hole with a nonlinear electric source}}, \href{https://doi.org/10.1007/s10773-015-2521-6}{\emph{Int. J. Theor. Phys.} {\bfseries 54} (2015) 2855} [\href{https://arxiv.org/abs/1408.3334}{{\ttfamily 1408.3334}}].

\bibitem{Ghosh:2014pba}
S.G.~Ghosh, \emph{{A nonsingular rotating black hole}}, \href{https://doi.org/10.1140/epjc/s10052-015-3740-y}{\emph{Eur. Phys. J. C} {\bfseries 75} (2015) 532} [\href{https://arxiv.org/abs/1408.5668}{{\ttfamily 1408.5668}}].

\bibitem{Simpson:2019mud}
A.~Simpson and M.~Visser, \emph{{Regular black holes with asymptotically Minkowski cores}}, \href{https://doi.org/10.3390/universe6010008}{\emph{Universe} {\bfseries 6} (2019) 8} [\href{https://arxiv.org/abs/1911.01020}{{\ttfamily 1911.01020}}].

\bibitem{Simpson:2018tsi}
A.~Simpson and M.~Visser, \emph{{Black-bounce to traversable wormhole}}, \href{https://doi.org/10.1088/1475-7516/2019/02/042}{\emph{JCAP} {\bfseries 02} (2019) 042} [\href{https://arxiv.org/abs/1812.07114}{{\ttfamily 1812.07114}}].

\bibitem{Peltola:2008pa}
A.~Peltola and G.~Kunstatter, \emph{{A Complete, Single-Horizon Quantum Corrected Black Hole Spacetime}}, \href{https://doi.org/10.1103/PhysRevD.79.061501}{\emph{Phys. Rev. D} {\bfseries 79} (2009) 061501} [\href{https://arxiv.org/abs/0811.3240}{{\ttfamily 0811.3240}}].

\bibitem{Peltola:2009jm}
A.~Peltola and G.~Kunstatter, \emph{{Effective Polymer Dynamics of D-Dimensional Black Hole Interiors}}, \href{https://doi.org/10.1103/PhysRevD.80.044031}{\emph{Phys. Rev. D} {\bfseries 80} (2009) 044031} [\href{https://arxiv.org/abs/0902.1746}{{\ttfamily 0902.1746}}].

\bibitem{Bianchi:2018mml}
E.~Bianchi, M.~Christodoulou, F.~D'Ambrosio, H.M.~Haggard and C.~Rovelli, \emph{{White Holes as Remnants: A Surprising Scenario for the End of a Black Hole}}, \href{https://doi.org/10.1088/1361-6382/aae550}{\emph{Class. Quant. Grav.} {\bfseries 35} (2018) 225003} [\href{https://arxiv.org/abs/1802.04264}{{\ttfamily 1802.04264}}].

\bibitem{DAmbrosio:2018wgv}
F.~D'Ambrosio and C.~Rovelli, \emph{{How information crosses Schwarzschild{\textquoteright}s central singularity}}, \href{https://doi.org/10.1088/1361-6382/aae499}{\emph{Class. Quant. Grav.} {\bfseries 35} (2018) 215010} [\href{https://arxiv.org/abs/1803.05015}{{\ttfamily 1803.05015}}].

\bibitem{Zhang:2024khj}
C.~Zhang, J.~Lewandowski, Y.~Ma and J.~Yang, \emph{{Black holes and covariance in effective quantum gravity}}, \href{https://doi.org/10.1103/PhysRevD.111.L081504}{\emph{Phys. Rev. D} {\bfseries 111} (2025) L081504} [\href{https://arxiv.org/abs/2407.10168}{{\ttfamily 2407.10168}}].

\bibitem{Bueno:2024dgm}
P.~Bueno, P.A.~Cano and R.A.~Hennigar, \emph{{Regular black holes from pure gravity}}, \href{https://doi.org/10.1016/j.physletb.2025.139260}{\emph{Phys. Lett. B} {\bfseries 861} (2025) 139260} [\href{https://arxiv.org/abs/2403.04827}{{\ttfamily 2403.04827}}].

\bibitem{Bueno:2024eig}
P.~Bueno, P.A.~Cano, R.A.~Hennigar and {\'A}.J.~Murcia, \emph{{Dynamical Formation of Regular Black Holes}}, \href{https://doi.org/10.1103/PhysRevLett.134.181401}{\emph{Phys. Rev. Lett.} {\bfseries 134} (2025) 181401} [\href{https://arxiv.org/abs/2412.02742}{{\ttfamily 2412.02742}}].

\bibitem{Bueno:2024zsx}
P.~Bueno, P.A.~Cano, R.A.~Hennigar and {\'A}.J.~Murcia, \emph{{Regular black holes from thin-shell collapse}}, \href{https://doi.org/10.1103/PhysRevD.111.104009}{\emph{Phys. Rev. D} {\bfseries 111} (2025) 104009} [\href{https://arxiv.org/abs/2412.02740}{{\ttfamily 2412.02740}}].

\bibitem{Bueno:2025zaj}
P.~Bueno, P.A.~Cano, R.A.~Hennigar and {\'A}.J.~Murcia, \emph{{Regular black hole formation in four-dimensional non-polynomial gravities}},  \href{https://arxiv.org/abs/2509.19016}{{\ttfamily 2509.19016}}.

\bibitem{Vertogradov:2025snh}
V.~Vertogradov, A.~{\"O}vg{\"u}n and D.~Shatov, \emph{{Formation of regular black hole from baryonic matter*}}, \href{https://doi.org/10.1088/1674-1137/ade95c}{\emph{Chin. Phys.} {\bfseries 49} (2025) 115103} [\href{https://arxiv.org/abs/2502.00521}{{\ttfamily 2502.00521}}].

\bibitem{Dialektopoulos:2025fzp}
K.F.~Dialektopoulos, T.~Papanikolaou and V.~Zarikas, \emph{{Primordial black hole-driven cosmic acceleration}}, \href{https://doi.org/10.1142/S0218271825440158}{\emph{Int. J. Mod. Phys. D} {\bfseries 34} (2025) 2544015} [\href{https://arxiv.org/abs/2511.18080}{{\ttfamily 2511.18080}}].

\bibitem{Dialektopoulos:2025mfz}
K.~Dialektopoulos, T.~Papanikolaou and V.~Zarikas, \emph{{Primordial black holes as cosmic expansion accelerators}}, \href{https://doi.org/10.1016/j.physletb.2025.139948}{\emph{Phys. Lett. B} {\bfseries 870} (2025) 139948} [\href{https://arxiv.org/abs/2502.18352}{{\ttfamily 2502.18352}}].

\bibitem{Cadoni:2023lum}
M.~Cadoni, A.P.~Sanna, M.~Pitzalis, B.~Banerjee, R.~Murgia, N.~Hazra et~al., \emph{{Cosmological coupling of nonsingular black holes}}, \href{https://doi.org/10.1088/1475-7516/2023/11/007}{\emph{JCAP} {\bfseries 11} (2023) 007} [\href{https://arxiv.org/abs/2306.11588}{{\ttfamily 2306.11588}}].

\bibitem{Farrah:2023opk}
D.~Farrah et~al., \emph{{Observational Evidence for Cosmological Coupling of Black Holes and its Implications for an Astrophysical Source of Dark Energy}}, \href{https://doi.org/10.3847/2041-8213/acb704}{\emph{Astrophys. J. Lett.} {\bfseries 944} (2023) L31} [\href{https://arxiv.org/abs/2302.07878}{{\ttfamily 2302.07878}}].

\bibitem{Croker:2024jfg}
K.S.~Croker, G.~Tarl{\'e}, S.P.~Ahlen, B.G.~Cartwright, D.~Farrah, N.~Fernandez et~al., \emph{{DESI dark energy time evolution is recovered by cosmologically coupled black holes}}, \href{https://doi.org/10.1088/1475-7516/2024/10/094}{\emph{JCAP} {\bfseries 10} (2024) 094} [\href{https://arxiv.org/abs/2405.12282}{{\ttfamily 2405.12282}}].

\bibitem{Faraoni:2024ghi}
V.~Faraoni and M.~Rinaldi, \emph{{Black hole event horizons are cosmologically coupled}}, \href{https://doi.org/10.1103/PhysRevD.110.063553}{\emph{Phys. Rev. D} {\bfseries 110} (2024) 063553} [\href{https://arxiv.org/abs/2407.14549}{{\ttfamily 2407.14549}}].

\bibitem{Calza:2024qxn}
M.~Calz{\`a}, F.~Gianesello, M.~Rinaldi and S.~Vagnozzi, \emph{{Implications of cosmologically coupled black holes for pulsar timing arrays}}, \href{https://doi.org/10.1038/s41598-024-82661-8}{\emph{Sci. Rep.} {\bfseries 14} (2024) 31296} [\href{https://arxiv.org/abs/2409.01801}{{\ttfamily 2409.01801}}].

\bibitem{Davies:2024ysj}
P.C.W.~Davies, D.A.~Easson and P.B.~Levin, \emph{{Nonsingular black holes as dark matter}}, \href{https://doi.org/10.1103/PhysRevD.111.103512}{\emph{Phys. Rev. D} {\bfseries 111} (2025) 103512} [\href{https://arxiv.org/abs/2410.21577}{{\ttfamily 2410.21577}}].

\bibitem{Trivedi:2025vry}
O.~Trivedi and A.~Loeb, \emph{{Could planck star remnants be dark matter?}}, \href{https://doi.org/10.1016/j.dark.2025.102003}{\emph{Phys. Dark Univ.} {\bfseries 49} (2025) 102003} [\href{https://arxiv.org/abs/2506.03334}{{\ttfamily 2506.03334}}].

\bibitem{Ong:2024dnr}
Y.C.~Ong, \emph{{The Case For Black Hole Remnants: A Review}},  11, 2024 [\href{https://arxiv.org/abs/2412.00322}{{\ttfamily 2412.00322}}].

\bibitem{Dymnikova:2015yma}
I.~Dymnikova and M.~Khlopov, \emph{{Regular black hole remnants and graviatoms with de Sitter interior as heavy dark matter candidates probing inhomogeneity of early universe}}, \href{https://doi.org/10.1142/S0218271815450029}{\emph{Int. J. Mod. Phys. D} {\bfseries 24} (2015) 1545002} [\href{https://arxiv.org/abs/1510.01351}{{\ttfamily 1510.01351}}].

\bibitem{Khodadi:2025icd}
M.~Khodadi, \emph{{Could regular primordial black holes be dark matter?}}, \href{https://doi.org/10.1016/j.physletb.2025.139922}{\emph{Phys. Lett. B} {\bfseries 870} (2025) 139922} [\href{https://arxiv.org/abs/2509.20483}{{\ttfamily 2509.20483}}].

\bibitem{Calza:2024fzo}
M.~Calz{\`a}, D.~Pedrotti and S.~Vagnozzi, \emph{{Primordial regular black holes as all the dark matter. I. Time-radial-symmetric metrics}}, \href{https://doi.org/10.1103/PhysRevD.111.024009}{\emph{Phys. Rev. D} {\bfseries 111} (2025) 024009} [\href{https://arxiv.org/abs/2409.02804}{{\ttfamily 2409.02804}}].

\bibitem{Calza:2024xdh}
M.~Calz{\`a}, D.~Pedrotti and S.~Vagnozzi, \emph{{Primordial regular black holes as all the dark matter. II. Non-time-radial-symmetric and loop quantum gravity-inspired metrics}}, \href{https://doi.org/10.1103/PhysRevD.111.024010}{\emph{Phys. Rev. D} {\bfseries 111} (2025) 024010} [\href{https://arxiv.org/abs/2409.02807}{{\ttfamily 2409.02807}}].

\bibitem{Calza:2025mwn}
M.~Calz{\`a}, D.~Pedrotti, G.-W.~Yuan and S.~Vagnozzi, \emph{{Primordial regular black holes as all the dark matter. III. Covariant canonical quantum gravity models}},  \href{https://arxiv.org/abs/2507.02396}{{\ttfamily 2507.02396}}.

\bibitem{Loc:2025mzc}
N.P.D.~Loc, \emph{{Dark matter production from evaporation of regular primordial black holes}}, \href{https://doi.org/10.1140/epjc/s10052-026-15811-z}{\emph{Eur. Phys. J. C} {\bfseries 86} (2026) 559} [\href{https://arxiv.org/abs/2510.09395}{{\ttfamily 2510.09395}}].

\bibitem{Loc:2025whx}
N.P.D.~Loc, \emph{{Constraining regular primordial black holes with isocurvature gravitational waves}},  \href{https://arxiv.org/abs/2512.04548}{{\ttfamily 2512.04548}}.

\bibitem{Vagnozzi:2022moj}
S.~Vagnozzi et~al., \emph{{Horizon-scale tests of gravity theories and fundamental physics from the Event Horizon Telescope image of Sagittarius A}}, \href{https://doi.org/10.1088/1361-6382/acd97b}{\emph{Class. Quant. Grav.} {\bfseries 40} (2023) 165007} [\href{https://arxiv.org/abs/2205.07787}{{\ttfamily 2205.07787}}].

\bibitem{Flachi:2012nv}
A.~Flachi and J.P.S.~Lemos, \emph{{Quasinormal modes of regular black holes}}, \href{https://doi.org/10.1103/PhysRevD.87.024034}{\emph{Phys. Rev. D} {\bfseries 87} (2013) 024034} [\href{https://arxiv.org/abs/1211.6212}{{\ttfamily 1211.6212}}].

\bibitem{Pedrotti:2024znu}
D.~Pedrotti and S.~Vagnozzi, \emph{{Quasinormal modes-shadow correspondence for rotating regular black holes}}, \href{https://doi.org/10.1103/PhysRevD.110.084075}{\emph{Phys. Rev. D} {\bfseries 110} (2024) 084075} [\href{https://arxiv.org/abs/2404.07589}{{\ttfamily 2404.07589}}].

\bibitem{Cardoso:2016rao}
V.~Cardoso, E.~Franzin and P.~Pani, \emph{{Is the gravitational-wave ringdown a probe of the event horizon?}}, \href{https://doi.org/10.1103/PhysRevLett.116.171101}{\emph{Phys. Rev. Lett.} {\bfseries 116} (2016) 171101} [\href{https://arxiv.org/abs/1602.07309}{{\ttfamily 1602.07309}}].

\bibitem{Rosato:2026apq}
R.F.~Rosato, F.~Crescimbeni, S.~Yi, E.~Berti and P.~Pani, \emph{{Novel Ringdown Tests of General Relativity with Black Hole Graybody Factors}}, \href{https://doi.org/10.1103/37g5-r3k8}{\emph{Phys. Rev. Lett.} {\bfseries 137} (2026) 121402} [\href{https://arxiv.org/abs/2604.11895}{{\ttfamily 2604.11895}}].

\bibitem{Lu:2025vol}
N.~Lu, S.~Ma, O.J.~Piccinni, Y.~Chen and L.~Sun, \emph{{GW250114 reveals signatures of post-merger black-hole horizon}}, \href{https://doi.org/10.1038/s41586-026-10696-0}{\emph{Nature} {\bfseries 655} (2026) 71} [\href{https://arxiv.org/abs/2510.01001}{{\ttfamily 2510.01001}}].

\bibitem{Krishnendu:2025rud}
N.~Krishnendu, V. et~al., \emph{{Implications of GW241011 for Rotating Exotic Compact Objects}}, \href{https://doi.org/10.1103/29y5-nx9y}{\emph{Phys. Rev. Lett.} {\bfseries 137} (2026) 071402} [\href{https://arxiv.org/abs/2511.17341}{{\ttfamily 2511.17341}}].

\bibitem{Wang:2025oek}
R.~Wang, Q.-L.~Shi, W.~Xiong and P.-C.~Li, \emph{{Tidal Love numbers for regular black holes}},  \href{https://arxiv.org/abs/2512.05767}{{\ttfamily 2512.05767}}.

\bibitem{Andres-Carcasona:2025bni}
M.~Andr{\'e}s-Carcasona and G.~Caneva~Santoro, \emph{{No Love for black holes: tightest constraints on tidal Love numbers of black holes from GW250114}},  \href{https://arxiv.org/abs/2512.01918}{{\ttfamily 2512.01918}}.

\bibitem{Rao:2025rop}
X.-P.~Rao and H.~Huang, \emph{{Can we distinguish whether black holes have singularities or not through echoes and light rings?}},  \href{https://arxiv.org/abs/2505.11073}{{\ttfamily 2505.11073}}.

\bibitem{Akbarieh:2023kjv}
A.R.~Akbarieh, M.~Khoshragbaf and M.~Atazadeh, \emph{{Accretion disk around regular black holes}}, \href{https://doi.org/10.1142/S0219887824503237}{\emph{Int. J. Geom. Meth. Mod. Phys.} {\bfseries 22} (2025) 2450323} [\href{https://arxiv.org/abs/2302.02784}{{\ttfamily 2302.02784}}].

\bibitem{Boshkayev:2023fft}
K.~Boshkayev, T.~Konysbayev, Y.~Kurmanov, O.~Luongo, M.~Muccino, A.~Taukenova et~al., \emph{{Luminosity of accretion disks around rotating regular black holes}}, \href{https://doi.org/10.1140/epjc/s10052-024-12446-w}{\emph{Eur. Phys. J. C} {\bfseries 84} (2024) 230} [\href{https://arxiv.org/abs/2307.15003}{{\ttfamily 2307.15003}}].

\bibitem{Page:1976df}
D.N.~Page, \emph{{Particle Emission Rates from a Black Hole: Massless Particles from an Uncharged, Nonrotating Hole}}, \href{https://doi.org/10.1103/PhysRevD.13.198}{\emph{Phys. Rev. D} {\bfseries 13} (1976) 198}.

\bibitem{Page:1976ki}
D.N.~Page, \emph{{Particle Emission Rates from a Black Hole. 2. Massless Particles from a Rotating Hole}}, \href{https://doi.org/10.1103/PhysRevD.14.3260}{\emph{Phys. Rev. D} {\bfseries 14} (1976) 3260}.

\bibitem{Dong:2015yjs}
R.~Dong, W.H.~Kinney and D.~Stojkovic, \emph{{Gravitational wave production by Hawking radiation from rotating primordial black holes}}, \href{https://doi.org/10.1088/1475-7516/2016/10/034}{\emph{JCAP} {\bfseries 10} (2016) 034} [\href{https://arxiv.org/abs/1511.05642}{{\ttfamily 1511.05642}}].

\bibitem{Ireland:2023avg}
A.~Ireland, S.~Profumo and J.~Scharnhorst, \emph{{Primordial gravitational waves from black hole evaporation in standard and nonstandard cosmologies}}, \href{https://doi.org/10.1103/PhysRevD.107.104021}{\emph{Phys. Rev. D} {\bfseries 107} (2023) 104021} [\href{https://arxiv.org/abs/2302.10188}{{\ttfamily 2302.10188}}].

\bibitem{Anantua:2008am}
R.~Anantua, R.~Easther and J.T.~Giblin, \emph{{GUT-Scale Primordial Black Holes: Consequences and Constraints}}, \href{https://doi.org/10.1103/PhysRevLett.103.111303}{\emph{Phys. Rev. Lett.} {\bfseries 103} (2009) 111303} [\href{https://arxiv.org/abs/0812.0825}{{\ttfamily 0812.0825}}].

\bibitem{Dolgov:2011cq}
A.D.~Dolgov and D.~Ejlli, \emph{{Relic gravitational waves from light primordial black holes}}, \href{https://doi.org/10.1103/PhysRevD.84.024028}{\emph{Phys. Rev. D} {\bfseries 84} (2011) 024028} [\href{https://arxiv.org/abs/1105.2303}{{\ttfamily 1105.2303}}].

\bibitem{Phinney:2001di}
E.S.~Phinney, \emph{{A Practical theorem on gravitational wave backgrounds}},  \href{https://arxiv.org/abs/astro-ph/0108028}{{\ttfamily astro-ph/0108028}}.

\bibitem{Sasaki:2018dmp}
M.~Sasaki, T.~Suyama, T.~Tanaka and S.~Yokoyama, \emph{{Primordial black holes{\textemdash}perspectives in gravitational wave astronomy}}, \href{https://doi.org/10.1088/1361-6382/aaa7b4}{\emph{Class. Quant. Grav.} {\bfseries 35} (2018) 063001} [\href{https://arxiv.org/abs/1801.05235}{{\ttfamily 1801.05235}}].

\bibitem{Wang:2019kaf}
S.~Wang, T.~Terada and K.~Kohri, \emph{{Prospective constraints on the primordial black hole abundance from the stochastic gravitational-wave backgrounds produced by coalescing events and curvature perturbations}}, \href{https://doi.org/10.1103/PhysRevD.99.103531}{\emph{Phys. Rev. D} {\bfseries 99} (2019) 103531} [\href{https://arxiv.org/abs/1903.05924}{{\ttfamily 1903.05924}}].

\bibitem{Sasaki:2016jop}
M.~Sasaki, T.~Suyama, T.~Tanaka and S.~Yokoyama, \emph{{Primordial Black Hole Scenario for the Gravitational-Wave Event GW150914}}, \href{https://doi.org/10.1103/PhysRevLett.117.061101}{\emph{Phys. Rev. Lett.} {\bfseries 117} (2016) 061101} [\href{https://arxiv.org/abs/1603.08338}{{\ttfamily 1603.08338}}].

\bibitem{Ajith:2007kx}
P.~Ajith et~al., \emph{{A Template bank for gravitational waveforms from coalescing binary black holes. I. Non-spinning binaries}}, \href{https://doi.org/10.1103/PhysRevD.77.104017}{\emph{Phys. Rev. D} {\bfseries 77} (2008) 104017} [\href{https://arxiv.org/abs/0710.2335}{{\ttfamily 0710.2335}}].

\bibitem{Zhu:2011bd}
X.-J.~Zhu, E.~Howell, T.~Regimbau, D.~Blair and Z.-H.~Zhu, \emph{{Stochastic Gravitational Wave Background from Coalescing Binary Black Holes}}, \href{https://doi.org/10.1088/0004-637X/739/2/86}{\emph{Astrophys. J.} {\bfseries 739} (2011) 86} [\href{https://arxiv.org/abs/1104.3565}{{\ttfamily 1104.3565}}].

\bibitem{Ali-Haimoud:2017rtz}
Y.~Ali-Ha{\"\i}moud, E.D.~Kovetz and M.~Kamionkowski, \emph{{Merger rate of primordial black-hole binaries}}, \href{https://doi.org/10.1103/PhysRevD.96.123523}{\emph{Phys. Rev. D} {\bfseries 96} (2017) 123523} [\href{https://arxiv.org/abs/1709.06576}{{\ttfamily 1709.06576}}].

\bibitem{Raidal:2024bmm}
M.~Raidal, V.~Vaskonen and H.~Veerm{\"a}e, \emph{{Formation of~Primordial Black Hole Binaries and~Their Merger Rates}},  in \emph{{Primordial Black Holes}}, C.~Byrnes, G.~Franciolini, T.~Harada, P.~Pani and M.~Sasaki, eds. (2025), \href{https://doi.org/10.1007/978-981-97-8887-3_16}{DOI} [\href{https://arxiv.org/abs/2404.08416}{{\ttfamily 2404.08416}}].

\bibitem{Raidal:2017mfl}
M.~Raidal, V.~Vaskonen and H.~Veerm{\"a}e, \emph{{Gravitational Waves from Primordial Black Hole Mergers}}, \href{https://doi.org/10.1088/1475-7516/2017/09/037}{\emph{JCAP} {\bfseries 09} (2017) 037} [\href{https://arxiv.org/abs/1707.01480}{{\ttfamily 1707.01480}}].

\bibitem{LIGOScientific:2025kry}
{\scshape LIGO Scientific, VIRGO, KAGRA} collaboration, \emph{{Cosmological and High Energy Physics implications from gravitational-wave background searches in LIGO-Virgo-KAGRA's O1-O4a runs}},  \href{https://arxiv.org/abs/2510.26848}{{\ttfamily 2510.26848}}.

\bibitem{Hawking:1974rv}
S.W.~Hawking, \emph{{Black hole explosions}}, \href{https://doi.org/10.1038/248030a0}{\emph{Nature} {\bfseries 248} (1974) 30}.

\bibitem{Hawking:1975vcx}
S.W.~Hawking, \emph{{Particle Creation by Black Holes}}, \href{https://doi.org/10.1007/BF02345020}{\emph{Commun. Math. Phys.} {\bfseries 43} (1975) 199}.

\bibitem{Gibbons:1977mu}
G.W.~Gibbons and S.W.~Hawking, \emph{{Cosmological Event Horizons, Thermodynamics, and Particle Creation}}, \href{https://doi.org/10.1103/PhysRevD.15.2738}{\emph{Phys. Rev. D} {\bfseries 15} (1977) 2738}.

\bibitem{LIGOScientific:2016aoc}
{\scshape LIGO Scientific, Virgo} collaboration, \emph{{Observation of Gravitational Waves from a Binary Black Hole Merger}}, \href{https://doi.org/10.1103/PhysRevLett.116.061102}{\emph{Phys. Rev. Lett.} {\bfseries 116} (2016) 061102} [\href{https://arxiv.org/abs/1602.03837}{{\ttfamily 1602.03837}}].

\bibitem{LIGOScientific:2016wyt}
{\scshape LIGO Scientific, Virgo} collaboration, \emph{{The basic physics of the binary black hole merger GW150914}}, \href{https://doi.org/10.1002/andp.201600209}{\emph{Annalen Phys.} {\bfseries 529} (2017) 1600209} [\href{https://arxiv.org/abs/1608.01940}{{\ttfamily 1608.01940}}].

\bibitem{Planck:2018vyg}
{\scshape Planck} collaboration, \emph{{Planck 2018 results. VI. Cosmological parameters}}, \href{https://doi.org/10.1051/0004-6361/201833910}{\emph{Astron. Astrophys.} {\bfseries 641} (2020) A6} [\href{https://arxiv.org/abs/1807.06209}{{\ttfamily 1807.06209}}].

\bibitem{Inomata:2020lmk}
K.~Inomata, M.~Kawasaki, K.~Mukaida, T.~Terada and T.T.~Yanagida, \emph{{Gravitational Wave Production right after a Primordial Black Hole Evaporation}}, \href{https://doi.org/10.1103/PhysRevD.101.123533}{\emph{Phys. Rev. D} {\bfseries 101} (2020) 123533} [\href{https://arxiv.org/abs/2003.10455}{{\ttfamily 2003.10455}}].

\bibitem{Hooper:2019gtx}
D.~Hooper, G.~Krnjaic and S.D.~McDermott, \emph{{Dark Radiation and Superheavy Dark Matter from Black Hole Domination}}, \href{https://doi.org/10.1007/JHEP08(2019)001}{\emph{JHEP} {\bfseries 08} (2019) 001} [\href{https://arxiv.org/abs/1905.01301}{{\ttfamily 1905.01301}}].

\bibitem{Hooper:2020evu}
D.~Hooper, G.~Krnjaic, J.~March-Russell, S.D.~McDermott and R.~Petrossian-Byrne, \emph{{Hot Gravitons and Gravitational Waves From Kerr Black Holes in the Early Universe}},  \href{https://arxiv.org/abs/2004.00618}{{\ttfamily 2004.00618}}.

\bibitem{Papanikolaou:2020qtd}
T.~Papanikolaou, V.~Vennin and D.~Langlois, \emph{{Gravitational waves from a universe filled with primordial black holes}}, \href{https://doi.org/10.1088/1475-7516/2021/03/053}{\emph{JCAP} {\bfseries 03} (2021) 053} [\href{https://arxiv.org/abs/2010.11573}{{\ttfamily 2010.11573}}].

\bibitem{Domenech:2020ssp}
G.~Dom\`enech, C.~Lin and M.~Sasaki, \emph{{Gravitational wave constraints on the primordial black hole dominated early universe}}, \href{https://doi.org/10.1088/1475-7516/2021/11/E01}{\emph{JCAP} {\bfseries 04} (2021) 062} [\href{https://arxiv.org/abs/2012.08151}{{\ttfamily 2012.08151}}].

\bibitem{Carr:2020gox}
B.~Carr, K.~Kohri, Y.~Sendouda and J.~Yokoyama, \emph{{Constraints on primordial black holes}}, \href{https://doi.org/10.1088/1361-6633/ac1e31}{\emph{Rept. Prog. Phys.} {\bfseries 84} (2021) 116902} [\href{https://arxiv.org/abs/2002.12778}{{\ttfamily 2002.12778}}].

\bibitem{Kumar:2020bqf}
A.~Kumar, D.V.~Singh and S.G.~Ghosh, \emph{{Hayward black holes in Einstein{\textendash}Gauss{\textendash}Bonnet gravity}}, \href{https://doi.org/10.1016/j.aop.2020.168214}{\emph{Annals Phys.} {\bfseries 419} (2020) 168214} [\href{https://arxiv.org/abs/2003.14016}{{\ttfamily 2003.14016}}].

\bibitem{Addazi:2021xuf}
A.~Addazi et~al., \emph{{Quantum gravity phenomenology at the dawn of the multi-messenger era{\textemdash}A review}}, \href{https://doi.org/10.1016/j.ppnp.2022.103948}{\emph{Prog. Part. Nucl. Phys.} {\bfseries 125} (2022) 103948} [\href{https://arxiv.org/abs/2111.05659}{{\ttfamily 2111.05659}}].

\bibitem{Kinney:2005vj}
W.H.~Kinney, \emph{{Horizon crossing and inflation with large eta}}, \href{https://doi.org/10.1103/PhysRevD.72.023515}{\emph{Phys. Rev. D} {\bfseries 72} (2005) 023515} [\href{https://arxiv.org/abs/gr-qc/0503017}{{\ttfamily gr-qc/0503017}}].

\bibitem{Biagetti:2021eep}
M.~Biagetti, V.~De~Luca, G.~Franciolini, A.~Kehagias and A.~Riotto, \emph{{The formation probability of primordial black holes}}, \href{https://doi.org/10.1016/j.physletb.2021.136602}{\emph{Phys. Lett. B} {\bfseries 820} (2021) 136602} [\href{https://arxiv.org/abs/2105.07810}{{\ttfamily 2105.07810}}].

\bibitem{Carr:1975qj}
B.J.~Carr, \emph{{The Primordial black hole mass spectrum}}, \href{https://doi.org/10.1086/153853}{\emph{Astrophys. J.} {\bfseries 201} (1975) 1}.

\bibitem{Kohri:2018qtx}
K.~Kohri and T.~Terada, \emph{{Primordial Black Hole Dark Matter and LIGO/Virgo Merger Rate from Inflation with Running Spectral Indices: Formation in the Matter- and/or Radiation-Dominated Universe}}, \href{https://doi.org/10.1088/1361-6382/aaea18}{\emph{Class. Quant. Grav.} {\bfseries 35} (2018) 235017} [\href{https://arxiv.org/abs/1802.06785}{{\ttfamily 1802.06785}}].

\bibitem{Kohri:2018awv}
K.~Kohri and T.~Terada, \emph{{Semianalytic calculation of gravitational wave spectrum nonlinearly induced from primordial curvature perturbations}}, \href{https://doi.org/10.1103/PhysRevD.97.123532}{\emph{Phys. Rev. D} {\bfseries 97} (2018) 123532} [\href{https://arxiv.org/abs/1804.08577}{{\ttfamily 1804.08577}}].

\bibitem{Domenech:2021ztg}
G.~Dom{\`e}nech, \emph{{Scalar Induced Gravitational Waves Review}}, \href{https://doi.org/10.3390/universe7110398}{\emph{Universe} {\bfseries 7} (2021) 398} [\href{https://arxiv.org/abs/2109.01398}{{\ttfamily 2109.01398}}].

\bibitem{Loc:2024qbz}
N.P.D.~Loc, \emph{{Gravitational waves from burdened primordial black holes dark matter}}, \href{https://doi.org/10.1103/PhysRevD.111.023509}{\emph{Phys. Rev. D} {\bfseries 111} (2025) 023509} [\href{https://arxiv.org/abs/2410.17544}{{\ttfamily 2410.17544}}].

\bibitem{Caprini:2015zlo}
C.~Caprini et~al., \emph{{Science with the space-based interferometer eLISA. II: Gravitational waves from cosmological phase transitions}}, \href{https://doi.org/10.1088/1475-7516/2016/04/001}{\emph{JCAP} {\bfseries 04} (2016) 001} [\href{https://arxiv.org/abs/1512.06239}{{\ttfamily 1512.06239}}].

\bibitem{Inomata:2019ivs}
K.~Inomata, K.~Kohri, T.~Nakama and T.~Terada, \emph{{Enhancement of Gravitational Waves Induced by Scalar Perturbations due to a Sudden Transition from an Early Matter Era to the Radiation Era}}, \href{https://doi.org/10.1103/PhysRevD.108.049901}{\emph{Phys. Rev. D} {\bfseries 100} (2019) 043532} [\href{https://arxiv.org/abs/1904.12879}{{\ttfamily 1904.12879}}].

\bibitem{Inomata:2025wiv}
K.~Inomata, K.~Kohri and T.~Terada, \emph{{The poltergeist mechanism -- Enhancement of scalar-induced gravitational waves with early matter-dominated era}},  \href{https://arxiv.org/abs/2511.07266}{{\ttfamily 2511.07266}}.

\bibitem{Allahverdi:2025btx}
R.~Allahverdi, J.B.~Dent, N.P.D.~Loc and T.~Xu, \emph{{Gravitational wave signatures of primordial black hole accretion during early matter domination}}, \href{https://doi.org/10.1088/1475-7516/2025/10/026}{\emph{JCAP} {\bfseries 10} (2025) 026} [\href{https://arxiv.org/abs/2505.21885}{{\ttfamily 2505.21885}}].

\bibitem{Carroll:2004st}
S.M.~Carroll, \emph{{Spacetime and Geometry}: {An Introduction to General Relativity}}, Cambridge University Press (7, 2019), \href{https://doi.org/10.1017/9781108770385}{10.1017/9781108770385}.

\bibitem{Cen:2025atm}
Y.~Cen and Y.~Song, \emph{{Upper bound on the radius of the innermost stable circular orbit of black holes}}, \href{https://doi.org/10.1016/j.physletb.2025.139545}{\emph{Phys. Lett. B} {\bfseries 866} (2025) 139545} [\href{https://arxiv.org/abs/2505.02107}{{\ttfamily 2505.02107}}].

\bibitem{Planck:2018jri}
{\scshape Planck} collaboration, \emph{{Planck 2018 results. X. Constraints on inflation}}, \href{https://doi.org/10.1051/0004-6361/201833887}{\emph{Astron. Astrophys.} {\bfseries 641} (2020) A10} [\href{https://arxiv.org/abs/1807.06211}{{\ttfamily 1807.06211}}].

\bibitem{Galanti:2021bxt}
G.~Galanti and M.~Roncadelli, \emph{{Precision cosmology made more precise}},  \href{https://arxiv.org/abs/2102.01637}{{\ttfamily 2102.01637}}.

\bibitem{Choi:2024acs}
K.-Y.~Choi, E.~Lkhagvadorj and S.~Mahapatra, \emph{{Gravitational wave sourced by decay of massive particle from primordial black hole evaporation}}, \href{https://doi.org/10.1088/1475-7516/2024/07/064}{\emph{JCAP} {\bfseries 07} (2024) 064} [\href{https://arxiv.org/abs/2403.15269}{{\ttfamily 2403.15269}}].

\end{thebibliography}\endgroup

\end{document}